%% file: draft.tex
\documentclass[hidelinks,12pt]{article}
\usepackage[letterpaper,margin=1.25in]{geometry}

\usepackage{setspace}
\usepackage{pgfplots}
\usepgfplotslibrary{groupplots,fillbetween}
\pgfplotsset{compat=1.18}
\usepackage[T1]{fontenc}
\usepackage[utf8]{inputenc}

\usepackage[sc]{mathpazo}

\usepackage{amsmath,amssymb,amsthm,mathtools}
\usepackage{microtype}
\usepackage{enumitem}
\usepackage{hyperref}
\usepackage{natbib}

\newtheorem{theorem}{Theorem}
\newtheorem{proposition}{Proposition}
\newtheorem{corollary}{Corollary}

\newcommand{\CS}{\mathrm{CS}}
\newcommand{\CSmin}{\underline{\Delta{\mathrm{CS}}}}
\newcommand{\CSmax}{\overline{\Delta{\mathrm{CS}}}}

\title{\textbf{Price Discrimination and Revealed Consumer Welfare}\footnote{We thank Ali Shourideh, Arthur Taburet, Doron Ravid and Rahul Deb for helpful discussions.}
\vspace{1em} 
\\}
\author{ João Thereze\footnote{The Fuqua School of Business, Duke University--- \href{mailto:joao.thereze@duke.edu}{joao.thereze@duke.edu}.}  ~~~~~~~~~~~~ Udayan Vaidya\footnote{The Fuqua School of Business, Duke University--- \href{mailto:udayan.vaidya@duke.edu}{udayan.vaidya@duke.edu}.} }
\date{\today}

\begin{document}
\maketitle

\begin{abstract}
    We take a revealed-preference approach to study the welfare effects of price discrimination.
    An analyst observes price and quantity data before and after a monopolist segments the market. 
    The analyst knows neither the aggregate demand curve nor the information available to the monopolist. 
    We characterize the set of observable price-quantity pairs that are rationalizable by some market segmentation. 
    We then obtain sharp bounds on the change in consumer surplus arising from some segmentation. 
    Finally, we provide conditions under which the data identifies an unambiguous increase (decrease) in consumer surplus. 
\end{abstract}

\newpage 
\section{Introduction}

Firms are increasingly able to personalize prices using abundant consumer data and sophisticated prediction methods. 
Even as laws restricting personalized pricing enter legislative dockets, policy often lags practice. 
This creates a need for methods to study the effects of price discrimination using observed data, rather than relying on ex-ante predictions alone to shape policy. 
Such a retrospective approach could be used to evaluate the effects of implemented policies and guide enforcement action against existing practices. 
This empirical motivation guides our study of two related questions: when is observed market data consistent with price discrimination, and what can we conclude about the resulting consequences for consumer surplus?

In this paper, we take a revealed preference approach to studying the welfare effects of third-degree price discrimination. We consider an analyst who observes the price and quantity data generated by a revenue-maximizing monopolist over two periods. In the first period, all transactions happen at a uniform price, while in the second period, transaction prices vary. The analyst conjectures that the price changes reflect a market segmentation enabled by some signal about the consumers' willingness to pay. However, the analyst knows neither the underlying aggregate demand curve, nor the information available to the firm. Assuming the data was generated by market segmentation and the firm chose revenue-maximizing prices, what can the analyst infer from the observed data?

We begin by asking what observed data can be rationalized by price discrimination. 
That is, when does there exist an aggregate demand curve and corresponding segmentation such that the firm's prices maximize revenue in each period?
We show that rationalizability is characterized by $n+1$ simple inequalities, where $n$ is the number of price-quantity pairs observed in the second period. 
The first $n$ inequalities are consequences of optimality of the original uniform price.
To describe these conditions, first note that there is a minimal demand curve for each segment that is non-increasing and consistent with its observed price-quantity pair.
The inequalities require the original revenue to be greater than the hypothetical revenue generated by each observed price against the sum of these minimal demands.
The final inequality is a consequence of Blackwell's theorem: the firm's total revenue under price discrimination should be no less than with the uniform price. 

These $n+1$ conditions are also jointly sufficient to rationalize the observed data.
This is not immediate because rationalizability also requires each segment price to be optimal and for the segment demands to sum to the aggregate demand curve.
None of our $n+1$ inequalities checks these conditions directly.
For example, choosing the minimal segment demands can satisfy optimality, but will generally produce too little aggregate quantity at the original uniform price. 
However, increasing the demand to match the aggregate quantity risks violating price optimality.
The proof shows that this tension can be reduced to the choice of each segment's counterfactual demand at the uniform price. 
When the $n+1$ inequalities are jointly satisfied, there is enough flexibility to construct segment demands that simultaneously satisfy all of these properties.

We then study what changes in consumer surplus can be identified by data arising under price discrimination.
Our main results characterize the set of changes in consumer surplus that are consistent with the data. 
The set of achievable changes in surplus is an interval, and we obtain closed-form solutions for its upper and lower bounds.
When this set lies entirely on one side of zero, the data robustly reveal whether consumers gain or lose from price discrimination. We  provide simple sufficient conditions for each case, building on the closed-form characterization of the identified set.
Our conditions leverage optimality of the firm's prices and are therefore weaker than existing sufficient conditions relying on downward-sloping demand alone.

The forces that discipline our bounds are clearest in the special case where the analyst observes two prices in the second period, one above the uniform price and one below. 
For illustration, consider solving for the demands that maximize the gain in consumer surplus across periods.
Intuitively, we would like the demand curve corresponding to the low-price segment to be as high as possible because consumers in this segment are the ones who benefit from discrimination: either because they now buy at a lower price, or because they did not purchase at all under the uniform price. 
But this low-segment demand cannot be too flat; else, the firm would have incentive to raise the segment price.
At the same time, if the low-segment demand were too steep, it may violate optimality of the uniform price, since it may incentivize the firm to marginally reduce that price. 
The demand that maximizes consumer gains turns out to be the highest one respecting both of these revenue constraints in addition to consistency of the aggregate demand.

The same logic extends to characterizing welfare bounds when there are arbitrarily many segments. However, when there are multiple prices above and below the uniform price, distributing demand across segments to maximize consumer surplus becomes an additional allocation problem. For the lower bound, the data imply a minimum amount of demand that must remain in the low-price segments at the original price. Because each unit of demand generates a smaller consumer-surplus gain in segments whose prices are closer to the original price, minimizing the surplus gain requires allocating this mass greedily, beginning with those segments. The remaining part of the construction relies on bounding how large the demand can be between observed prices, which pinned down by the lower envelope of the price optimality constraints. An analogous argument characterizes the best-case rationalization. In both cases, these constructions yield closed-form expressions for the sharp identified set.

Our results provide a way to evaluate instances of price discrimination without identifying the underlying demand curves or market segmentation.
This is often a key challenge in empirical work, where constructing a counterfactual demand estimate may require substantial data and strong structural assumptions.
By contrast, the sharp identified set we characterize has a closed-form representation and is straightforward to compute from market data.
These bounds can therefore serve as a screening tool for regulators, helping to test when the available data are informative about welfare changes or when further structural analysis may be warranted.
Finally, because our rationalizability conditions can reject certain data as  inconsistent with price discrimination, they can also help distinguish price dispersion arising from market segmentation from dispersion due to  other forces such as learning, demand shifts, or cost fluctuations.

\paragraph{Related Literature} Our paper contributes to the classical literature on the welfare effects of monopolistic third-degree price discrimination. Since \citet{pigou1920economics} and \citet{robinson1933economics}, this literature has emphasized that discrimination affects welfare through both changes in total output and changes in the allocation of output across consumers. \citet{schmalensee1981output} and \citet{varian1985price} show that an increase in total output is necessary for third-degree price discrimination to increase total welfare. \citet{varian1985price} also derives welfare bounds for the case of two price segments, leveraging a single optimality condition that the firm must weakly benefit from segmentation. 
A subsequent literature sharpens these predictions by imposing additional restrictions on the shape of demand \citep{shih1988general,schwartz1990third,cowan2007welfare,aguirre2010monopoly,cowan2012third, cowan2016welfare}. Rather than imposing exogenous restrictions on the class of demands, our revealed-preference approach derives such restrictions from observed data and firm optimality alone. Moreover, we use all constraints imposed by optimality to obtain tight bounds on welfare changes.

We also complement a modern literature studying price discrimination through the lens of information design \citep{bergemann2015limits, haghpanah2022limits,haghpanah2023pareto,farboodi2025good,elliott2026market,kartik2026lemonade}. 
Closest to us is \citet{farboodi2025good}, who characterize properties of demand families such that any segmentation will unambiguously lead to a welfare loss or gain.
\citet{farboodi2025good} and the other papers in this literature typically take the underlying demand as known and ask which pairs of consumer and producer surplus are compatible with some information structure.
Our question is, in some ways, the inverse.
While they take demand curves as primitives, we take the observed data as primitive and are initially agnostic to which demand curves generate them.

Methodologically, we employ a revealed-preference approach, using finite observed data to discipline unobserved primitives. This technique has been applied successfully for individual optimization  \citep{afriat1967construction, deb2023revealed}, as well as general equilibrium \citep{brown1996testable} and imperfectly competitive markets \citep{carvajal2013revealed}. Like these papers, we see this nonparametric approach as complementary to the estimation strategies used in the empirical industrial organization literature.
Our rationalizability result characterizes when an underlying segmentation can generate observed prices and quantities, and our welfare results provide bounds that are robust to functional form assumptions. In that sense, we join a recent literature that studies robust tests for welfare analysis in different settings \citep{kang2025robustness,chassang2022robust}.

\section{Model} \label{sec:model}
We study the market for a single good provided by a monopolist with zero marginal costs.\footnote{Our results extend naturally if marginal costs are constant but different from zero.} There are two dates, $t\in\{0,1\}$. At the initial date $t=0$, the consumer's demand for the product is given by the demand function $D_0$. At $t=1$, the seller has additional data about consumers which allows her to price different segments of the market differently. In particular, at date $t=1$ the seller faces $n\geq2$ price segments with corresponding demand $D_j$, for $j \in \{1,\ldots,n\}$. We denote $[n] := \{0,...,n\}$.

The analyst observes the prices and quantities transacted in each market segment (and at time 0), given by the dataset $\mathbf{O}:=\{(p_j, q_j)\}_{j\in[n]}$, but not the demand functions that generated those outcomes. We assume, without loss of generality, that all price observations are distinct.\footnote{If two prices $p_i,p_j$ are the same, and $i,j\neq0$, then adding the quantities together allows us to collapse the two segments into an observationally equivalent one. If $p_i=p_0$, this observation does not affect consumer surplus. Moreover, this assumption does not play a role in the characterization of rationalizability.} 
What can the analyst say about the change of consumer surplus between the two dates? In particular, when can it be guaranteed that the market segmentation has benefited or hurt consumers?

To answer these questions formally, we begin by introducing some definitions.
 Fix a dataset $\mathbf{O}$. We will say that a vector of demand functions $\mathbf{D}:=\{D_j:\mathbb{R}\to\mathbb{R}_+\}_{j\in[n]}$ is \textbf{feasible} with respect to $\mathbf{O}$ if it satisfies the following conditions:
\begin{enumerate}
   \item \textbf{Consistency.} For each $j \in [n]$, $D_j(p_j) = q_j$, and $D_j$ is non-increasing ;
   \item \textbf{Aggregation.} $D_0 = \sum^n_{j=1} D_j$
   \item \textbf{Optimality.} $R_j := p_j q_j \geq p D_j(p)$, for all $j\in[n]$ and $\ p\geq 0.$
\end{enumerate}

Consistency requires that each $D_j$ satisfies the law of demand, and that prices and quantities are mutually consistent given the demand curve. Aggregation states that the segment demands add up to the aggregate demand, so demand did not change across the two periods. Finally, optimality guarantees that each price $p_j$ is revenue-maximizing with respect to demand $j$. 

Denote by $\mathcal{D}(\mathbf O)$ the set of all feasible demand functions for observable data $\mathbf{O}$. We say that a dataset $\mathbf{O}$ is \textbf{rationalizable} if the associated set of feasible demand functions $\mathcal{D}(\mathbf{O})$ is non-empty. That is, a set of observables is rationalizable if it can be understood as the optimal price-quantity pairs obtained by some market segmentation faced by the monopolist.

The analyst aims to investigate what changes in consumer surplus can be rationalized by the observed data $\mathbf{O}$.
Given a vector of demand functions $\mathbf{D}:= \{D_j\}_{j\in[n]}$, we define consumer surplus for each date $t\in\{0,1\}$ in the usual way:
\[
\CS_0 \left(\mathbf{D} \right) := \int^\infty_{p_0} D_0(p)dp, \qquad \text{and} \qquad \CS_1 \left( \mathbf{D} \right) := \sum^n_{j=1} \int^\infty_{p_j} D_j(p) dp. 
\]
\noindent
Denote the change in consumer surplus by $\Delta\CS (\mathbf{D}) := \CS_1 (\mathbf{D}) - \CS_0 (\mathbf{D})$. The bounds on the change in consumer surplus are then defined as

\[
\CSmax = \sup_{\mathbf{D}\in\mathcal{D}(\mathbf{O})} \Delta CS \qquad \text{and} \qquad \CSmin = \inf_{\mathbf{D}\in\mathcal{D}(\mathbf{O})} \Delta CS.
\]

Any consumer surplus change strictly between $\CSmin$ and $\CSmax$ is achievable by some demand vector because the set of feasible demands with respect to $\mathbf{O}$ is convex and  $\Delta \CS (\mathbf{D})$ is linear.
The main goal of our analysis is to characterize the bounds $\CSmin$ and $\CSmax$.

\section{Rationalizability}

In this section, we ask whether the dataset observed by the analyst is consistent with price discrimination: that is, whether $\mathbf{O}$ is rationalizable. Before answering whether price discrimination has benefited or harmed consumers, the analyst must first learn whether her observations are consistent with price discrimination to begin with. Therefore, characterizing rationalizability is a natural first step. Moreover, the characterization will provide insights about the nature of the revealed-preference problem.

We start by obtaining necessary conditions for rationalizability. First, note that the revenue in segment $j$, $R_j:=p_jq_j$, is observable to the analyst. Second, observe that the assumption of non-increasing, non-negative demands alone provides a simple lower bound for the demand function in each segment at date $t=1$, which we dub $\underline{D}_j$, for $j\geq1$. In particular, for any price lower than $p_j$, the demand in that segment cannot be lower than $q_j$ by monotonicity, and for any price larger than $p_j$, it cannot be below zero. Formally:

$$\underline{D}_j (p) := \mathbf{1}[p\leq p_j] q_j. $$

These lower-bounds in demand for each segment imply a lower bound for the aggregate demand in the market, which we denote $\underline{D}(p) := \sum^n_{j=1} \underline{D}_j (p).$

This lower bound allows us to derive a first set of necessary conditions for rationalizability. Concretely, optimality of the initial uniform price $p_0$ implies that $p_0$ must achieve a higher revenue than any other observed price against the lowest possible aggregate demand $\underline{D}$: 

\begin{equation}
    p_j \underline{D}(p_j) \leq R_0, \qquad j=1,\ldots,n.
    \label{eq:feas-p0optimality}
\end{equation}

Indeed, because the demand at each price is at least $\underline{D}$, the revenue obtained at that demand is the minimum revenue available for any aggregate demand that rationalizes $\mathbf{O}$. Therefore, if that inequality is violated, no set of demands consistent with the dataset can make $p_0$ optimal in the aggregate.

The second necessary condition is that, because the seller is better informed at $t=1$ than at $t=0$, he must obtain higher revenue by Blackwell's theorem:
\begin{equation}
R_0
\le
\sum_{i=1}^{\ell}R_i^L
+
\sum_{j=1}^hR_j^H.
\label{eq:feas-blackwell}
\end{equation}

The main result in this section proves that these conditions are also sufficient.

\begin{theorem}[Rationalizability] \label{thm:rationalizability}
A dataset $\mathbf{O}$ is rationalizable if and only if conditions \eqref{eq:feas-p0optimality} and \eqref{eq:feas-blackwell} hold.
\end{theorem}

Theorem~\ref{thm:rationalizability} states that, with $n+1$ observations, the observable implications of third-degree price discrimination are summarized by $n+1$ inequalities. $n$ of these inequalities, represented by condition \eqref{eq:feas-p0optimality}, assert that the original price $p_0$ must dominate any other observed price $p_j$ at the minimal demand allowed at that price. The remaining equation just observes that being able to segment the market cannot make the monopolist worse off. Remarkably, no extra conditions on the optimality of specific segments arise in this characterization. 

For an intuition on the result, consider the case of two segments, that is, $n=2$. Assume $p_1 < p_0<p_2$ and suppose that conditions \eqref{eq:feas-p0optimality} and \eqref{eq:feas-blackwell} hold. A natural candidate for a feasible demand vector is precisely the lower bound vector $\{\underline{D},\underline{D}_1,\underline{D}_2\}$. Each of those demands is consistent and, because $\underline{D}_j$ is a step function dropping to zero after the observed price $p_j$, each segment demand satisfies optimality. Moreover,  by Condition~\eqref{eq:feas-p0optimality}, the aggregate demand also does so. The only issue is that, because $\underline{D}$ is an underestimate, the demands might violate aggregation at $p_0$: $\underline{D}(p_0) < q_0$. 

To solve that issue, we increase the demands greedily. We start by modifying segment $j=1$, turning it into a two-step demand curve. Rather than drop to zero after price $p_1$, we progressively increase the minimal demand for prices between $(p_1, p_0]$. The new demand function, $\widetilde{D}_1$, equals $q_1$ up to $p_1$, takes a value $s_1>0$ for prices in $(p_1,p_0]$, and drops to zero thereafter. This can be done until either aggregation holds at $p_0$---in which case we are done---or $p_0$ becomes an optimal price for segment 1, that is $p_0 \tilde{D}_1(p_0) := p_0 s_1 = R_1$. If the latter happens, we fix $s_1 = R_1/p_0$ and start increasing the demand for segment 2 for prices between 0 and $p_0$. The only possible problem with this procedure would be if the increase in the demand of segment 2 needed to obtain aggregation, call that $s_2:= q_0-s_1$, was so high that $p_0$ becomes strictly optimal for segment 2. However, that is impossible under Condition~\eqref{eq:feas-blackwell}. Indeed:
$$R_0 \leq R_1 + R_2 = p_0 s_1 + p_0 s_2,$$ 
implying that $s_1+s_2 \geq q_0.$ 

The proof of Theorem~\ref{thm:rationalizability} works in the same way as the argument above. We construct a vector of segment demands that each take at most three values. Two of these values are pinned down by the observed quantity and zero. The third value is chosen through a greedy algorithm similar to the one above, pinning down each counterfactual demand $D_j(p_0)$ at the uniform price $p_0$.

\section{The Case of Two Segments}
We now move on to the goal of characterizing the maximal and minimal consumer surplus consistent with a given dataset. To provide intuition for the general result, we start from the simplest case in which $n=2$. That is, the seller divides the original population into two segments. Note that if $\max\{p_1,p_2\} \leq p_0$, it is clear that consumer surplus weakly increases, as all consumers face lower prices. The opposite can be inferred if $\min\{p_1,p_2\} \geq p_0$. Therefore, we will assume from now on that 

$$p^L := p_1 < p_0 < p_2 =: p^H, $$
and say that an outcome $\mathbf{O}$ that satisfies the inequality above is \textbf{non-trivial}.

Fix a set of feasible demands $\mathbf{D} =\{D_0,D^L,D^H\} $. Recalling that, by aggregation, $D^L+D^H=D_0$, we can rewrite the change in consumer surplus as:

\[
\Delta\CS(\mathbf{D}) = \int^{p_0}_{p^L} D^L(p)dp - \int^{p^H}_{p_0} D^H(p)dp.
\]

The first term above is the gain accrued by consumers who were allocated to the segment $L$ and, as a consequence, have experienced a reduction in prices. The second term is the loss, accrued by consumers who now face higher prices at $H$. Consider first the demand functions that minimize the consumer surplus gain. In the absence of feasibility constraints, the goal would be to make $D^H$ as high as possible (and $D^L$ as low as possible) in the relevant interval.

\subsection{The lower bound with two segments}

Naturally, feasibility imposes restrictions on the demand $D^H$. One such restriction is that optimality of $p^H$ in segment $H$ implies a maximal level of demand for any price:

\begin{equation}
R^H\geq p D^H(p). 
\label{eq: H-optimality}
\end{equation}

Additionally, optimality of $p_0$ in the aggregate 
demand puts a restriction on the level of the $H$-segment demand at that price:

\begin{equation}
R_0\geq p^L \left(q^L+D^H(p^L)\right) \geq p^L\left(q^L+D^H(p_0)\right).      
\label{eq: 0-optimality}
\end{equation}

Putting these two observations together provides an upper bound $\bar{q}^H$ for the level of $H$-demand at the original price, $p_0$, given by

\[D^H(p_0) \leq \min\left\{\frac{R^H}{p_0},\frac{R_0}{p^L}-q^L \right\} := \bar{q}^H.\] This inequality provides an upper bound, denoted by $\bar{q}^H$, to the mass of consumers in segment $H$ with willingness topay at least $p_0$. Our first main result essentially shows that the bounds imposed by $\bar{q}^H$ and optimality of $p^H$ interact to determine the worst rationalizable consumer surplus gain, $\CSmin.$

\begin{proposition}[$\CSmin$ with two segments]
\label{prop:lower-bound-two-segments}
Let $n=2$. Suppose $\mathbf{O}$ is rationalizable and non-trivial.
Then
\[
    \underline{\Delta CS}
    =
    (p_0-p^L)(q_0-\overline q^H)
    -
    \left[
        R^H-p_0\overline q^H
        +
        R^H
        \log\left(
            \frac{p^H\overline q^H}{R^H}
        \right)
    \right].
\]
\end{proposition}

The proof of Proposition~\ref{prop:lower-bound-two-segments} has two parts. We first prove that the expression in the statement is a lower bound for consumer surplus changes. This can be done by applying the upper bound for $D^H(p_0)$ we obtained in the text to bound both the gains and the losses in consumer surplus. Then, we explicitly describe the demand set $\mathbf{D}$ that achieves the lower bound. 

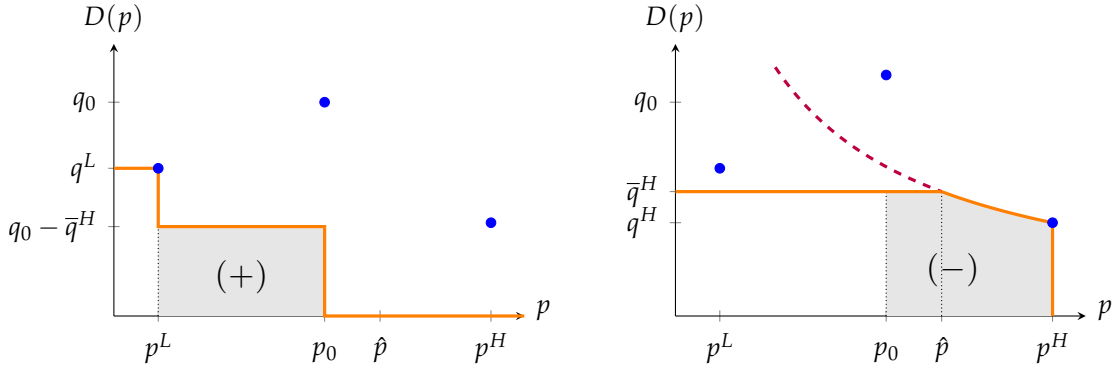
\begin{figure}[t!]
\centering
\input{Figures/worst-case-demand}
\caption{Worst-case demands.}
\label{fig:two-segment-lower-bound}
\end{figure}

Figure~\ref{fig:two-segment-lower-bound} illustrates the construction of these demand functions. In both panels, the blue dots represent the dataset, $\mathbf{O}$. Starting from the right panel, recall that, to maximize the losses, we want to find the highest feasible demand for segment $H$, $\underline{D}^H$ in the interval of prices $[p_0,p^H]$. This demand must however be consistent with the observations and optimal for observed prices. By inequality \eqref{eq: H-optimality}, optimality of $p^H$ provides an upper-bound for demand, which is given by the dashed purple curve in the graph. That upper-bound is the unit elastic demand which crosses the point $(p^H,q^H)$. In the absence of cross-segment constraints, the dashed curve would be the maximal demand compatible with the $H$ segment. 

However, optimality of the original price $p_0$  may preclude that high demand, as is the case in the picture: this is the content of inequality \eqref{eq: 0-optimality}. For an explanation, if the $H$ demand at price $p_L$ was the one indicated by the dashed purple curve, the seller would have incentives to lower prices from $p_0$ when facing the aggregate demand, which contradicts optimality of $p_0$. To make this deviation unprofitable, one must cap $\underline{D}^H$. Given that cap, the minimal $L$-demand consistent with $(p^L,q^L)$ is the step-function in the left panel. In particular, it cannot jump to $0$ at $p^L$ exactly because the aggregate demand is non-increasing and must be at least higher than $q_0$ at that price. 

Using the characterization in Proposition \ref{prop:lower-bound-two-segments}, we can provide a simple sufficient condition that guarantees consumers benefit from price discrimination.

\begin{corollary}
\label{cor:suff-reveal-gains}
    \[
    \frac{2p_0(p_0-p^L)}
    {(p^H+p_0)(p^H-p_0)}
    \ge
    \frac{q^H}{q_0-\overline q^H} \quad \implies \quad \CSmin > 0
    \]
\end{corollary}

This corollary follows by replacing negative term in Proposition \ref{prop:lower-bound-two-segments} with an upper bound, and rearranging the resulting expression. Graphically, this upper bound is formed by overestimating the gray negative region in Figure \ref{fig:two-segment-lower-bound} with a trapezoid that contains it. 

This expression has a natural interpretation. It says that the gain from cutting the price to the low price segment must be sufficiently large relative to the increased price in the high segment that is exposed to the price increase. Intuitively, this guarantees that the segmentation is sufficiently tilted towards expanding trade in the low segment rather than extracting surplus in a large, high segment. The condition is easier to satisfy when $p^L$ and $\overline{q}^H$ are small. 

\subsection{The upper bound with two segments}

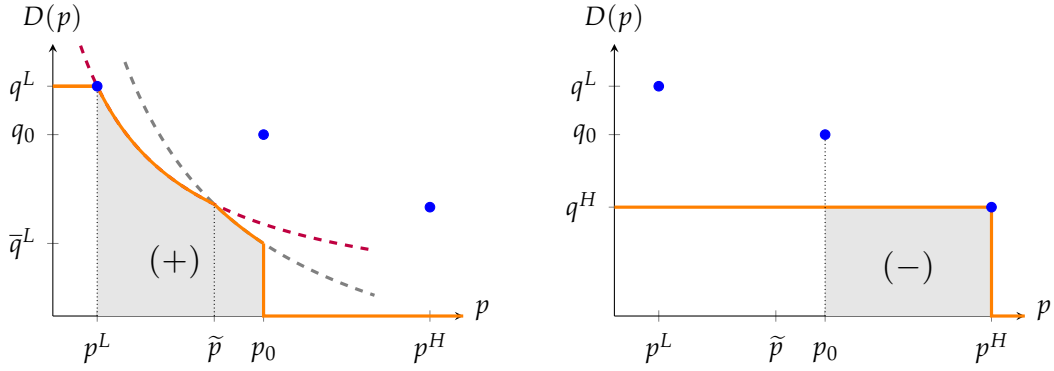
\begin{figure}[t]
\centering
\input{Figures/best-case-demand}
\caption{Best-case demands.}
\label{fig:two-segment-upper-bound}
\end{figure}

The construction of the upper bound for consumer surplus changes is similar to the one for the lower bound. To maximize consumer surplus, one would want to maximize the mass of people who enjoy the lower price, $p_1$. The ideal way to do so is to set $D^H$ to be the minimal possible demand consistent with the $(p^H, q^H)$, that is the step function on the right panel of Figure~\ref{fig:two-segment-upper-bound}. However, that imposes two constraints on $D^L$. 

The first constraint does not depend on $D^H$: the demand curve $D^L$ must lie below the iso-revenue curve crossing $(p^L,q^L)$:

$$p D^L(p) \leq R^L. $$
This constraint is given by dashed curve in purple on the left panel of Figure~\ref{fig:two-segment-upper-bound}. However, this constraint might not be sufficient. Indeed, given the demand function on the right, $p_0$ has to remain optimal for the aggregate market. Formally, for $p \in [p^L,p_0]$:

$$ p D^L(p) \leq R_0 - p q^H,   $$
which is the constraint given by the gray dashed curve on the left panel of Figure~\ref{fig:two-segment-upper-bound}. The demand curve for the low segment must then satisfy both of these constraints, giving rise to the full curve in that panel. 

Whether this pair of curves satisfies $L$-optimality depends on whether $D^L(p_0) = q_0 - q^H$ is low enough. When that is the case, the configuration in Figure~\ref{fig:two-segment-upper-bound} is optimal. Otherwise, more mass has to be given to the $H$ segment. Given these constructions, the upper bound in consumer surplus can be calculated as in the statement of the next proposition.

\begin{proposition}[$\CSmax$ with two segments]
\label{prop:upper-bound-two-segments}
Let $n=2$. Suppose $\mathbf{O}$ is rationalizable and non-trivial.
Then

\begin{equation*}
\label{eq:upper-bound}
    \overline{\Delta \CS}
    = \begin{cases} 
    R^L\log\frac{p_0}{p^L}
    -q^H(p^H-p_0)  & \text{if} \qquad R^L + p_0 q^H \leq R_0 \\ 
     R^L\log\frac{R_0-R^L}{p^L q^H}
    +R_0\log\frac{p_0 q^H}{R_0-R^L}
    -q^H(p^H-\frac{R_0-R^L}{q^H}) & \text{otherwise.}
    \end{cases}
\end{equation*}
\end{proposition}

Analogously to the case of revealing consumer gains, we can use this characterization to provide a simple sufficient condition under which the data reveal consumer losses.

\begin{corollary}
    \label{cor:suff-reveal-losses}
    \[
\frac{(p_0+p^L)(p_0-p^L)}
{2p_0(p^H-p_0)}
\leq
\frac{q^H}{q^L} \quad \implies  \quad \overline{\Delta \CS} < 0
\]
\end{corollary}

This corollary is derived in similar fashion to Corollary \ref{cor:suff-reveal-gains}, replacing the positive $\log$ term of $\overline{\Delta \CS}$ with a simple upper bound. Geometrically, this upper bound is formed by overestimating the  positive shaded region in Figure \ref{fig:two-segment-upper-bound} with a trapezoid that contains it. 

The interpretation of the condition in Corollary \ref{cor:suff-reveal-losses} is the inverse of Corollary \ref{cor:suff-reveal-gains}. This condition states that the high segment quantity is too large relative to the low segment quantity to outweigh the potential  price benefits. This condition is harder to satisfy when either the markup $(p^H - p_0)$ is small or the low segment quantity $q^L$ is large. 

\subsection{Examples of Revealed Gains and Losses}

We conclude this two-segment analysis with two examples of datasets that show how our characterization can be informative. The first example showcases observables in which consumers are sure to have gained from price discrimination. The second example, by changing only one of the data points, shows a situation in which consumers are known to have loss.

\paragraph{Example 1 (Revealing Consumer Gains).} Let $\mathbf{O}$ be such that 

\[ (p_0,q_0) = (1,1), \qquad (p^L,q^L)=(.8,.7), \qquad (p^H,q^H)=(1.2, .4). \]

In this case, 
$\bar{q}^H = \min\left\{\frac{1.2\times.4}{1}, \frac{1}{.8} - .7\right\} = \frac{1.2 \times .4}{1}=.48, $
implying the constraint on $D^H$ comes from optimality on the $H$-segment.

By 
Proposition~\ref{prop:lower-bound-two-segments}, we compute the minimum change in consumer surplus
$$\CSmin = (1-.8)(1-.48) - .48 \log\left(1.2\right) \approx.017 > 0. $$
As a consequence, it is safe for an analyst observing $\mathbf{O}$ to conclude, even without any knowledge of demand curves, that the observed price segmentation has increased consumer welfare. \hfill $\square$

\paragraph{Example 2 (Revealing Consumer Losses).} Now, assume $\mathbf{O'}$ is identical to $\mathbf{O}$, except that the high demand segment changes to

$$(p^H,q^H) = (2.4,.4). $$

Because $q^L+q^H = 1.1 > 1 = q_0$, it may not be clear that these observations are inconsistent with consumers being better off. Nevertheless, Proposition~\ref{prop:upper-bound-two-segments} proves that this is the case. Indeed, one can compute

$$\CSmax = .56 \log \left(\frac{1}{.8}\right) - 0.4 (2.4-1) \approx -.44 < 0. $$
\hfill $\square$

\section{General Bounds on Consumer Surplus}

The constructions for $n=2$ suggest that there is a qualitative difference in how our problem treats prices above and below the uniform price $p_0$. Indeed, it will be useful to partition the observations $\mathbf{O}/\{p_0,q_0\}$ into two groups:

$$L:= \{(p,q)\in\mathbf{O}: p<p_0 \}, \qquad H:=\{(p,q)\in\mathbf{O}:p>p_0\}.  $$

Let $|L|=\ell$ and $|H|=h$. We can then re-enumerate prices in $H$ and $L$ to write:
\[
p_1^L<\cdots<p_\ell^L<p_0<p_1^H<\cdots<p_h^H,
\]
and define $R^m_i = p^m_i q^m_i$, for $m\in\{L,H\}$, to be the revenue of the $i$-th segment in group $m$.

In this section we provide the counterparts of Propositions~\ref{prop:upper-bound-two-segments} and~\ref{prop:lower-bound-two-segments} for arbitrary segmentation. The formulas for $\CSmin$ and $\CSmax$ are unwieldy, but they are obtained in closed-form and trivial to compute. Moreover, the construction of the demand vectors that attain the bounds shares the same forces as the $n=2$ case. 

Our constructions start by obtaining adversarial bounds for the mass that must be allocated to the $L$ group. This is a lower bound, $\underline{Q}^L$, when we want to minimize consumer surplus change, and therefore want this group to be small. Conversely, it is an upper bound $\overline{Q}^L$ when we want to compute $\CSmax$. We then use similar inequalities to prove a bound for the consumer surplus changed that can be achieved. The proof is obtained by constructing demands that achieve the bound.

In the general case, however, there is an additional choice to be made: how to allocate mass across the segments within the groups. We prove that this should be done greedily starting from the segments closer to the price $p_0$.

We now state the results. To simplify notation, we define a clipping operator $]^A_{k}$, for $A=\{L,H\}$. For any number $b$, this operator is defined as:

$$ b]^A_k = \min\{p^A_{k+1}, \max\{p^A_k,b\}\}. $$

\begin{theorem}
\label{thm:lower-bound-general}
Suppose that $\mathbf{O}$ is rationalizable and define
\begin{equation}
\label{eq:Qlower}
\underline Q^L
=
\max\left\{
0,\,
q_0-\sum_{j=1}^h\frac{R_j^H}{p_0},\,
\max_{1\le k\le\ell}
\left[
q_0+\sum_{i=k}^{\ell}q_i^L-\frac{R_0}{p_k^L}
\right]
\right\}.
\end{equation}

For each
$k=0,\ldots,h-1$, define
\begin{equation}
\label{eq:pstar-high}
\hat{p}_k
=
 \left.\frac{
\displaystyle
\min\left\{
R_0,\,
\sum_{j=k+1}^h R_j^H
\right\}
}{
\displaystyle
\sum_{j=k+1}^h q_j^H
+
\min\left\{
q_0-\underline Q^L-\sum_{j=1}^h q_j^H,\,
\min_{1\le m\le k}
\left[
\frac{R_0}{p_m^H}
-\sum_{j=m}^h q_j^H
\right]
\right\}
}\right]_k^H,
\end{equation}
where the minimum over $m$ is interpreted as $+\infty$ when $k=0$. Then,
\begin{align}
\underline{\Delta CS}
={}&
\sum_{i=1}^{\ell}
(p_0-p_i^L)
\min\left\{
\frac{R_i^L}{p_0},
\left[
\underline Q^L
-
\sum_{a=i+1}^{\ell}\frac{R_a^L}{p_0}
\right]_+
\right\}
\nonumber\\
&-
\sum_{j=1}^h q_j^H(p_j^H-p_0)
-
\sum_{k=0}^{h-1}
\Bigg\{
\min\left\{
q_0-\underline Q^L-\sum_{j=1}^h q_j^H,\,
\min_{1\le r\le k}
\left[
\frac{R_0}{p_r^H}
-\sum_{j=r}^h q_j^H
\right]
\right\}
(\widehat p_k-p_k^H)
\nonumber\\
&\hspace{3.5cm}
+
\min\left\{
R_0,\,
\sum_{j=k+1}^hR_j^H
\right\}
\log\left(
\frac{p_{k+1}^H}{\widehat p_k}
\right)
\nonumber\\
&\hspace{3.5cm}
-
\left(
\sum_{j=k+1}^h q_j^H
\right)
(p_{k+1}^H-\widehat p_k)
\Bigg\}.
\label{eq:general-lower-bound}
\end{align}
\end{theorem}

Theorem~\ref{thm:lower-bound-general} characterizes the worst-case change in consumer surplus consistent with the observed data. There are two parts to the construction underlying Theorem ~\ref{thm:lower-bound-general}. First, the data impose a minimum amount of demand, $\underline{Q}^L$, that must be assigned across the low-price segments at the original price $p_0$. Because each unit of this demand generates a gain of at least $p_0-p_i^L$ in segment $i$, the worst-case allocation places prioritizes placing this mass in the low-price segments whose prices are closest to $p_0$. 

Second, conditional on this allocation, minimizing the gain in surplus involves making demand in the high-price segments as large as possible between $p_0$ and the observed high prices. However, this demand is constrained by optimality within each segment and  optimality of the original price in the aggregate market. The resulting upper envelope of these constraints determines the largest surplus loss that can be generated. The expression in~\eqref{eq:general-lower-bound} combines the smallest feasible gain from the low-price segments with this largest feasible loss from the high-price segments.

\begin{theorem}
\label{thm:upper-bound-general}
Suppose that $\mathbf{O}$ is rationalizable and define
\begin{equation}
\label{eq:Qupper}
\overline Q^L
=
\min\left\{
\sum_{i=1}^{\ell}\frac{R_i^L}{p_0},
\,
q_0-\sum_{j=1}^h q_j^H
\right\}.
\end{equation}

For notational convenience, let $p_{\ell+1}^L=p_0$. For each
$k=1,\ldots,\ell$, define
\begin{equation}
\tilde{p}_k
=
\left. \frac{
R_0-\sum_{i=1}^kR_i^L
}{
q_0-\overline Q^L+\sum_{i=k+1}^{\ell}q_i^L
}\right]^L_k,
\end{equation}
Then,
\begin{align}
\overline{\Delta CS}
={}&
\sum_{k=1}^{\ell}
\Bigg[
\left(\sum_{i=1}^kR_i^L\right)
\log\left(\frac{\widetilde p_k}{p_k^L}\right)
+
R_0
\log\left(\frac{p_{k+1}^L}{\widetilde p_k}\right)
\nonumber\\
&\hspace{2.5cm}
-
\left(
q_0-\overline Q^L
+
\sum_{i=k+1}^{\ell}q_i^L
\right)
(p_{k+1}^L-\widetilde p_k)
\Bigg]
\nonumber\\
&-
\sum_{j=1}^h q_j^H(p_j^H-p_0).
\label{eq:general-upper-bound}
\end{align}
\end{theorem}
Theorem~\ref{thm:upper-bound-general} gives the corresponding best-case change in consumer surplus. Here, the data imposes an upper bound, $\overline{Q}^L$, on the total demand that can be assigned to the low-price segments at $p_0$. 
As in the case with two segments, the total amount of mass to be distributed in each segment within group $L$ is constrained by two bounds: the one given by optimality of the segment, and the one obtained by aggregate optimality. Integrating the envelope of these two bounds across prices yields the maximal attainable gain on the low-price side. At the same time, consumer losses in the high-price segments are minimized by keeping those demands at their lowest levels consistent with the observed quantities.

\bibliographystyle{jpe.bst}
\bibliography{references}

\appendix
\section*{Appendix: Proofs}

\subsection*{Proof of Theorem~\ref{thm:rationalizability} }
Necessity is proved in the text. We show sufficiency by starting with a non-trivial $\mathbf{O}$ satisfying \eqref{eq:feas-p0optimality} and \eqref{eq:feas-blackwell} and constructing a set of demands that are feasible for $\mathbf{O}$. As we do later in the paper, we partition the observations $\mathbf{O}/\{p_0,q_0\}$ into two groups:

$$L:= \{(p,q)\in\mathbf{O}: p<p_0 \}, \qquad H:=\{(p,q)\in\mathbf{O}:p>p_0\}.  $$

We let $|L|=\ell$ and $|H|=h$, and  re-enumerate prices in $H$ and $L$ to write
\[
p_1^L<\cdots<p_\ell^L<p_0<p_1^H<\cdots<p_h^H.
\]
Finally, we define $R^m_i = p^m_i q^m_i$, for $m\in\{L,H\}$, to be the revenue of the $i$-th segment in group $m$.

We will construct the demands as a function of the variables $s^L_i$, $i=1,...,\ell$ and $s_j^H$, $j=1,...,h$, which we will define next.  For each $i =1,...,\ell$ the demand function is:
\[
D_i^L(p)
=
\begin{cases}
q_i^L, & p\le p_i^L,\\
s_i^L, & p_i^L<p\le p_0,\\
0, & p>p_0,
\end{cases}
\]
and for every $j=1,...,h$, define
\[
D_j^H(p)
=
\begin{cases}
s_j^H, & p\le p_0,\\
q_j^H, & p_0<p\le p_j^H,\\
0, & p>p_j^H.
\end{cases}
\]

For each $i=1,...,\ell$, define

$$u_i:=\frac{R_i^L}{p_0}, \qquad B_i:=q_0 + \sum^\ell_{k=i} q^L_k - \frac{R_0}{p^L_i}, \qquad S:= \max\left\{0,\max_{1\leq k\leq \ell } B_k, q_0 - \sum^h_{j=1}\frac{R^H_j}{p_0}\right\}. $$

We will later use these variables to define $s^L_i$ and $s^H_j$'s. We start by proving two technical results. We then define $s^L_i$ and $s^H_j$ and proceed to show that the demands satisfy the required conditions for rationalizability. 

\paragraph{Step 1. Bounds on $B_k$ and $S$.}

The goal of this step is to prove 
\begin{equation}
B_k\le\sum_{i=k}^{\ell}u_i
\qquad\text{for every }k=1,...,\ell,
\label{eq:B-capacity}
\end{equation}
and 
\begin{equation}
    S \leq \min\{\sum^\ell_{i=1} u_i, q_0-\sum^h_{j=1} q^H_j\}.
    \label{eq: upper-bound-S}
\end{equation} 

Start proving \eqref{eq:B-capacity}. For every $i \geq k$
\[
q_i^L-u_i
=
q_i^L\left(1-\frac{p_i^L}{p_0}\right)
\le
q_i^L\left(1-\frac{p_k^L}{p_0}\right).
\]
 Hence
\[
\begin{aligned}
\sum_{i=k}^{\ell}(q_i^L-u_i)
&\le
\left(1-\frac{p_k^L}{p_0}\right)
\sum_{i=k}^{\ell}q_i^L \leq \\ \left(1-\frac{p_k^L}{p_0}\right) \underline{D}(p_k) 
&\le
\left(1-\frac{p_k^L}{p_0}\right)
\frac{R_0}{p_k^L}\\
&=
\frac{R_0}{p_k^L}-q_0,
\end{aligned}
\]
where the second inequality follows from adding the non-negative quantities in $H$ to the summation and noting that the term in parenthesis is negative, and the third inequality follows from \eqref{eq:feas-p0optimality}. Rearranging gives
\eqref{eq:B-capacity}.

We shall now prove \eqref{eq: upper-bound-S}. First, we show $S\le\sum_{i=1}^{\ell}u_i.$ 
We show this for each term of $S$. The first term is trivial. For the terms $B_k$, this follows from \eqref{eq:B-capacity}; for the last
term, it follows from \eqref{eq:feas-blackwell} and the definition of $u_i$.

We now prove 
\begin{equation}
S\le q_0-\sum_{j=1}^h q_j^H.
\label{eq:S-upper}
\end{equation}
To see this, we prove that each item inside $S$ satisfies the inequality above. First, \eqref{eq:feas-p0optimality} implies, for the lowest price in group $H$
\[
\underline{D}(p^H_1) = \sum_{j=1}^h q_j^H
\le
\frac{R_0}{p_1^H}
<
q_0,
\]
so $q_0 - \sum^h_{j=1} q^H_j >0.$

Moreover, \eqref{eq:feas-p0optimality} also implies, for any price in $L$, $p^L_k$, $k=1,...,\ell$,
\[
B_k
=
q_0+\sum_{i=k}^{\ell}q_i^L-\frac{R_0}{p_k^L}
\le
q_0-\sum_{j=1}^h q_j^H
\]
for every $k$, implying that the second term in $S$ also satisfies the inequality. Finally,
\[
q_0-\sum_{j=1}^h\frac{R_j^H}{p_0}
\le
q_0-\sum_{j=1}^h q_j^H,
\]
where the second one from $p_j^H>p_0$ for all $j=1,...,h$.
This proves \eqref{eq:S-upper}.

\paragraph{Step 2. Feasibility of the demands in $L$.} We can finally define $s^L_\ell : = \min\{S,u_\ell\}$, and 
\begin{equation}
s_i^L
:=
\min\left\{
S,\sum^\ell_{k=i} u_k
\right\}-\min\left\{
S,\sum^\ell_{k=i+1} u_k
\right\},
\qquad i=1,...,\ell-1.
\label{eq:low-greedy}
\end{equation}

Because $u_i$'s are nonnegative, it is clear that $s^L_i \geq 0$, so $D^L_i$ is non-negative. We shall show that $s^L_i \leq u_i = \frac{R^L_i}{p_0} < \frac{R^L_i}{p^L_i}=q^L_i$, so $D^L_i$ is non-increasing. The second inequality follows from $p^L_i<p_0$. To prove the first inequality, note that if $S\leq \sum^\ell_{k=i+1} u_k$, then $s^L_i=0,$ and the inequality holds. Now, assume $S>\sum^\ell_{k=i+1}u_k$, then:

$$s^L_i = \min\left\{
S,\sum^\ell_{k=i} u_k
\right\}-\sum^\ell_{k=i+1}u_k \leq \sum^\ell_{k=i} u_k - \sum^\ell_{k=i+1}u_k = u_i, $$
proving the assertion. Therefore, $D^L_i$ is nonnegative and nonincreasing. It clearly is consistent at $p^L_i$. To conclude that $p^L_i$ is monopolist-optimal, notice first that for $p\leq p^L_i$ the demand is constant, so no $p<p^L_i$ can be optimal. Similarly, for $p>p_0$ demand is zero, so $p^L_i$ dominates any such price. So let $p^L_i<p\leq p_0$, to obtain

$$p D^L_i(p) = p s_i^L \leq p u_i = \frac{p}{p_0} R^L_i \leq R_i^L,$$
where the first inequality follows from $s^L_i \leq u_i$, as we proved in the last paragraph, and the second inequality follows from $p\leq p_0$. We therefore conclude that $p^L_i$ is optimal in segment $i$ for all $i=1,...,\ell$.

\paragraph{Step 3. Feasibility of Demands in $H$.} Note that 
\[
\sum_{j=1}^{\ell}s_j^L= \min\{S,\sum^\ell_{k=1}u_k\}= S,
\]
where the second equality follows from \eqref{eq: upper-bound-S}. The remaining mass at $p_0$ is then $q_0-S$. By \eqref{eq:S-upper},
\[
q_0-S\ge\sum_{j=1}^h q_j^H.
\]
By the definition of $S$,
\[
q_0-S
\le
\sum_{j=1}^h\frac{R_j^H}{p_0}.
\]
Hence there exist quantities $s_j^H\in
\left[
q_j^H,\frac{R_j^H}{p_0}
\right]$, $j=1,\ldots,h$, 
such that $\sum_{j=1}^h s_j^H=q_0-S.$ Choose any such set of $s^H_j$'s. By definition of such $s^H_j$'s, $D^H_j$ is non-negative and non-increasing. It is also clearly consistent with $p^H_j$.  To prove optimality in segment $j$, note that no $p \in (p_0,p^H_j)$ can be optimal because demand is constant in that interval and continuous at $p^H_j$. Moreover, no higher price than $p^H_j$ can be optimal because demand is zero there. So take $p \leq p_0$ to obtain:

$$p D^H_j = p s^H_j \leq \frac{p}{p_0} R^H_j \leq R^H_j, $$
where the first inequality is from the definition of $s^H_j$, and the second one by $p\leq p_0$.

\paragraph{Step 3. Feasibility of aggregate demand.}
We first rule out optimality of $p< p_0$. For $k=1,\ldots,\ell$, let $p$
\[
p_{k-1}^L<p\le p_k^L,
\]
where $p_0^L:=0$. Then, if $i=k,\ldots,\ell$, $D^L_i(p)=q^L_i$, whereas if $i<k$, $D^L_i=s^L_i$. Therefore,

\[
D_0(p) = \sum^{k-1}_{i=1} s_i^L +   \sum^\ell_{i=k} q^L_i + \sum^h_{j=1} s^H_j =  q_0+\sum_{i=k}^{\ell}(q_i^L-s_i^L),
\]
where the second equality follows from $\sum^h_{j=1} s^H_j = q_0 - S = q_0 - \sum^L_{i=1} s^L_i$.

Now, note $\sum_{i=k}^{\ell}s_i^L = \min\{S, \sum^\ell_{i=k} u_i\}\ge B_k$, because $S \geq B_k$ by the definition of $S$, and $\sum^\ell_{i=k} u_i \geq B_k$ by \eqref{eq:B-capacity}. Therefore, 
\[
D_0(p) = q_0+\sum_{i=k}^{\ell}(q_i^L-s_i^L)
\le
\frac{R_0}{p_k^L},
\]
and because $p\le p_k^L$,
\[
p
\left(
\sum_iD_i^L(p)+\sum_jD_j^H(p)
\right)
\le R_0.
\]
For $p_\ell^L<p\le p_0$, aggregate demand equals $q_0$, so the same conclusion follows directly from $p\le p_0$.

Now consider $p>p_0$. For $k=1,\ldots,h$, if
\[
p_{k-1}^H<p\le p_k^H,
\qquad p_0^H:=p_0,
\]
then aggregate demand satisfies
\[
D_0(p) = \sum_{j=k}^h q_j^H = \underline{D}(p) \leq \frac{R_0}{p^H_k},
\]
where the inequality follows by \eqref{eq:feas-p0optimality}. Therefore, by $p\leq p_k^H$
\[
p
\left(
\sum_iD_i^L(p)+\sum_jD_j^H(p)
\right)
\le R_0.
\]
Above $p_h^H$, aggregate demand is zero.

Thus $p_0$ maximizes aggregate revenue, each observed segment price
maximizes revenue in its segment, and all observed price-quantity pairs are
reproduced, proving $\mathbf{O}$ is rationalizable. \hfill $\blacksquare$

\subsection*{Proof of Proposition~\ref{prop:lower-bound-two-segments}}
Consider any rationalization of the observed outcomes. As established in the text:
\begin{equation}
    D^H(p_0)\le\overline q^H.
    \label{eq:two-segment-high-at-p0}
\end{equation}
We use this inequality to bound both the gains of buyers in segment $L$ and the losses of buyers in segment $H$.

\paragraph{Step 1. Bounding the gains in $L$.}
At $p_0$:
\[
    D^L(p_0)
    =
    q_0-D^H(p_0) \geq q_0 - \overline q^H,
\]
where the equality follows from aggregation of segments, and the inequality follows from \eqref{eq:two-segment-high-at-p0}.

Since demand is nonincreasing:

\begin{equation}
    \int_{p^L}^{p_0}D^L(p)\,dp
    \ge     \int_{p^L}^{p_0}D^L(p_0)\,dp \geq
    (p_0-p^L)(q_0-\overline q^H).
    \label{eq:two-segment-minimum-gain}
\end{equation}

\paragraph{Step 2. Bounding the losses in $H$.} Optimality of $p^H$ for $D^H$ implies $D^H(p)\le\frac{R^H}{p}$. Together with monotonicity of $D^H$, we obtain: 
\begin{equation}
    D^H(p)
    \le
    \min\left\{
        \overline q^H,\,
        \frac{R^H}{p}
    \right\},
    \qquad
    p\in[p_0,p^H].
    \label{eq:two-segment-high-envelope}
\end{equation}

Rationalizability implies
\[
    q^H\le\overline q^H\le\frac{R^H}{p_0}.
\]
Indeed, the first inequality follows because, as previously shown, $\overline{q}^H \geq D^H(p_0)$, and $q^H\leq D^H(p_0)$ by monotonicity; while the second equality is
immediate from the definition of $\overline{q}^H$. Consequently,  $\min\{\bar{q}_H, \frac{R^H}{p_0}\}=\bar{q}^H$, and $\min\{\bar{q}_H, \frac{R^H}{p^H}\}=q^H$, implying 
\[
    p_0\le\frac{R^H}{\overline{q}^H}\le p^H.
\]
Thus, integrating \eqref{eq:two-segment-high-envelope} obtains
\[
\begin{aligned}
    \int_{p_0}^{p^H}D^H(p)\,dp
    &\le
    \overline q^H\left(\frac{R^H}{\overline{q}^H}-p_0\right)
    +
    R^H\int_{\frac{R^H}{\overline{q}^H}}^{p^H}\frac{1}{p}\,dp\\
    &=
    R^H-p_0\overline q^H
    +
    R^H
    \log\left(
        \frac{p^H\overline q^H}{R^H}
    \right).
\end{aligned}
\]
Combining this inequality with
\eqref{eq:two-segment-minimum-gain} yields the expression for $\CSmin$ stated in the theorem. It remains to prove that the lower bound is attainable.

\paragraph{Step 3. Constructing $\mathbf{D}$ to achieve the lower bound.}

Let $\frac{R_0-R_1}{q_2}:=\frac{R^H}{\overline q^H}$, and define
\[
    \widehat D^L(p)
    :=
    \begin{cases}
        q^L,
        & p\le p^L,\\[4pt]
        q_0-\overline q^H,
        & p^L<p\le p_0,\\[4pt]
        0,
        & p>p_0,
    \end{cases}
\]
\[
    \widehat D^H(p)
    :=
    \begin{cases}
        \overline q^H,
        & p\le\frac{R_0-R_1}{q_2},\\[4pt]
        \dfrac{R^H}{p},
        & \frac{R_0-R_1}{q_2}<p\le p^H,\\[8pt]
        0,
        & p>p^H,
    \end{cases}
\]
and $\hat{D}_0 = \hat{D}^L+\hat{D}^H$.

We first verify that these demands are feasible given $\mathbf{O}$. First, because  $\mathbf{O}$ is rationalizable,
\[
   0 \leq q_0-\overline q^H\le q^L,
\]
so $\widehat D^L$ is non-increasing and non-negative. For the first inequality, note that rationalizability implies, for any feasible demand set $\mathbf{D}$:
$$R^H = p^HD^H(p^H) \leq p^H\left(D^H(p^H)+D^L(p^H)\right) \leq R_0, $$

and therefore $\bar{q}^H \leq \frac{R^H}{p_0}\leq q_0$.  

To prove the second inequality, we will show that each term in $\overline{q}^H$ is larger than $q_0-q^L$. For the first term, note that rationalizability implies
\[
    R_0\le R^L+R^H,
\]
and hence
\[
    \frac{R^H}{p_0}
    \ge
    q_0-\frac{R^L}{p_0}
    >
    q_0-q^L.
\]

For the second term, 
$$R_0 - R^L \geq p^L q_0 - R^L = p^L \left(q_0-q^L \right), $$
which shows that the second term also satisfies the inequality. Thus
\[
    \overline q^H\ge q_0-q^L.
\]

We now verify that $\widehat D^H$ is non-increasing and non-negative. Non-negativity is trivial.  Moreover, Step 2 established that rationalizability implies $\frac{R^H}{p}<\bar{q}^H$ if and only if $p \geq \hat{p}$, proving monotonicity. Clearly, $\widehat D^H(p^H)
    =
    \frac{R^H}{p^H}
    =
    q^H,$ establishing price-quantity consistency. 

We next verify optimality for each segment. For the high-price segment, by construction and choice of $\hat{p}$, $p\widehat D^H(p)\le R^H$ for every $p$, with equality for
$p\in[\frac{R_0-R_1}{q_2},p^H]$.

For the low-price segment, because $\widehat D^L$ is a step function with jumps at $p^L$ and $p_0$, these are the only two candidates for optimal price. We want to show $p_0 (q_0 - \bar{q}^H) \leq R^L$ or, equivalently
\[
    \overline q^H
    \ge
    q_0-\frac{R^L}{p_0}.
\]
To see that the inequality hold we prove that it holds for each term in $\bar{q}^H$. For the first term,$R^H/p_0$, it follows from $R_0\le R^L+R^H$, which is a consequence of rationalizability; and, since rationalizability also
implies $R_0\ge R^L$,
\[
    \frac{R_0-R^L}{p^L}
    \ge
    \frac{R_0-R^L}{p_0}
    =
    q_0-\frac{R^L}{p_0},
\]
which proves it for the second term. Thus $p^L$ is optimal in $L$.

It remains to verify aggregate optimality of $p_0$. For
$p^L<p\le p_0$, aggregate demand is exactly $q_0$, so
\[
    p\bigl(\widehat D^L(p)+\widehat D^H(p)\bigr)
    =
    pq_0
    \le R_0.
\]
For $p\le p^L$, aggregate demand is $q^L+\overline q^H$, and the definition
of $\overline q^H$ gives
\[
    p^L(q^L+\overline q^H)
    =
    R^L+p^L\overline q^H
    \le R_0.
\]
Therefore no price below $p_0$ yields aggregate revenue above $R_0$.

For $p>p_0$, the low-price demand is zero and
\[
    p\widehat D^H(p)\le R^H \leq R_0,
\]
by rationalizability. Hence no price above $p_0$ yields aggregate revenue above $R_0$.

Thus $\widehat D^L$ and $\widehat D^H$ constitute a valid rationalization. Under this
rationalization, the inequalities in
\eqref{eq:two-segment-minimum-gain} and
\eqref{eq:two-segment-high-envelope} binds. Therefore the resulting change in consumer surplus is
exactly the lower bound stated in the theorem. \hfill $\blacksquare$

\medskip

\subsection*{Proof of Proposition~\ref{prop:upper-bound-two-segments}}
Consider any rationalization of the observed outcomes. We first derive an
upper bound on the gain in the low-price segment and a lower bound on the loss
in the high-price segment.

\paragraph{Step 1. Bounding the gains in $L$.}
For every $p\in[p^L,p_0]$,
\[
    R_0 \ge p\bigl(D^L(p)+D^H(p)\bigr)\ge p \left(D^L(p)+q^H \right),
\]
where the first inequality follows from aggregate optimality of $p_0$, and the second from monotonicity of $D^H$. Reorganizing this inequality, and recalling that optimality in $L$ implies
\[
    D^L(p)\le\frac{R^L}{p},
\]
we obtain the following upper bound: 
\begin{equation}
    D^L(p)
    \le
    \min\left\{
        \frac{R^L}{p},\,
        \frac{R_0}{p}-q^H
    \right\},
    \qquad p\in[p^L,p_0].
\end{equation}

It follows that the gain in consumer surplus from $L$ satisfies
\begin{equation}
   \int_{p^L}^{p_0}D^L(p)\,dp \leq    \int_{p^L}^{p_0}
    \min\left\{
        \frac{R^L}{p},\,
        \frac{R_0}{p}-q^H
    \right\}\,dp.   
    \label{eq:two-segment-upper-low-envelope}
\end{equation}

\paragraph{Step 2. Bounding the losses in $H$.} For the consumer surplus loss from $H$, consider the bound:
  \[ \int_{p_0}^{p^H}D^H(p)\,dp \geq q^H (p^H-p_0),
  \]
  where the inequality follows from monotonicity of $D^H$ and price-quantity consistency at $p^H$.

Therefore every rationalization satisfies
\begin{equation}
    \Delta CS
    \le
    \int_{p^L}^{p_0}
    \min\left\{
        \frac{R^L}{p},\,
        \frac{R_0}{p}-q^H
    \right\}\,dp
    -
    q^H(p^H-p_0).   
    \label{eq: upper-bound-cs-two-segments}
\end{equation}

 We next show this upper bound is attainable.
\paragraph{Step 3. Constructing $\mathbf{D}$ to achieve the upper bound.}
Define
\[
    \overline q^L:=
    \min\left\{
        \frac{R^L}{p_0},\,
        q_0-q^H
    \right\}.
\]
Consider the following functions
\[
    \widehat D^L(p)
    :=
    \begin{cases}
        q^L,
        & p\le p^L,\\[4pt]
          \min\left\{
        \frac{R^L}{p},\,
        \frac{R_0}{p}-q^H
    \right\},
        & p^L<p\le p_0,\\[4pt]
        0,
        & p>p_0,
    \end{cases}
\]
\[
    \widehat D^H(p)
    :=
    \begin{cases}
        q_0-\overline q^L,
        & p\le p_0,\\[4pt]
        q^H,
        & p_0<p\le p^H,\\[4pt]
        0,
        & p>p^H.
    \end{cases}
\]
and $\widehat{D}_0 = \widehat D^L+\widehat D^H$. We first verify that these demands are feasible.

First, $\widehat D^L(p^L)=q^L$, so it is consistent. Moreover, by rationalizability of $\mathbf{O}$, there exists a feasible demand set $\mathbf{D}$, implying, for any $p \leq p_H$:

\[
p q^H \leq p D^H(p) \leq p \left(D^H(p) + D^L(p)\right) \leq R_0,
\]
where the first inequality follows from monotonicity of $D^H$, the second by non-negativity of $D^L$, and the third by optimality of $p_0$ in the aggregate. Reorganizing this inequality, one obtains that $\frac{R_0}{p}-q^H\geq 0$ for $p \leq p_0$. Therefore, $\hat{D}^L$ is non-negative. Given that it is non-negative, it is clearly non-increasing.

The $H$ demand is also clearly consistent. Moreover, $q_0-\bar{q}_L = \max\{q_0 - \frac{R^L}{p_0}, q^H\}\geq q^H \geq 0$, so $\widehat D^H$ is both non-negative and non-increasing. 

We next verify optimality, starting with $L$. We know that at $p^L$ the seller achieves $R^L$. Moreover, because $\widehat D^L$ is flat until $p^L$, no $p<p^L$ can be optimal. 
For $p^L<p\le p_0$,
\[
    p\widehat D^L(p) = p \min\{\frac{R^L}{p}, \frac{R_0}{p}-q^H\}\le R^L.
\]
Above $p_0$, demand is zero. Thus $p^L$ is revenue-maximizing.

For segment $H$, $\widehat{D}^H$ is a step function, so the two candidate optimal prices are $p_0$ and $p^H$. We know that at $p^H$ the seller obtains $R^H$. We claim that
\[
    q_0-\overline q^L\le\frac{R^H}{p_0}.
\]
Indeed, if $\overline q^L=q_0-q^H$, then the claim follows from
\[
    q^H<\frac{R^H}{p_0},
\]
since $p^H>p_0$. Otherwise, $\overline q^L=\frac{R^L}{p_0}$. Rationalizability implies $R_0\le R^L+R^H$, and hence
\[
    q_0-\frac{R^L}{p_0}
    \le
    \frac{R^H}{p_0},
\]
proving the claim.

It follows that, for $p\le p_0$,
\[
    p\widehat D^H(p)
    \le
    p_0(q_0-\overline q^L)
    \le
    R^H,
\]
and $p^H$ is profit maximizing.

It remains to verify aggregate optimality of $p_0$. For $p \leq p^L$, both demands are constant, so no $p<p^L$ can be optimal. For
$p\in[p^L,p_0]$,
\[
    \widehat D^L(p)=  \min\left\{         \frac{R^L}{p},\,         \frac{R_0}{p}-q^H     \right\}.
\]
There are two cases.

If $\overline q^L=q_0-q^H$, then $\widehat D^H(p)=q^H$ for $p\le p_0.$ Then
\[
    p\bigl(\widehat D^L(p)+\widehat D^H(p)\bigr) \le p\left( \frac{R_0}{p}-q^H+q^H \right)= R_0.
\]

For the second case, suppose instead that $\overline q^L=\frac{R^L}{p_0}\leq q_0-q^H.$ Then $\widehat D^H(p)
    =
    q_0-\frac{R^L}{p_0}$, for $p\le p_0.$
Since
\[
      \min\left\{         \frac{R^L}{p},\,         \frac{R_0}{p}-q^H     \right\}\le\frac{R^L}{p},
\]
aggregate demand satisfies
\[
\begin{aligned}
    \widehat D^L(p)+\widehat D^H(p)
    &\le
    \frac{R^L}{p}
    +
    q_0-\frac{R^L}{p_0}.
\end{aligned}
\]
Multiplying by $p$,
\[
\begin{aligned}
    p\bigl(\widehat D^L(p)+\widehat D^H(p)\bigr)
    &\le p\left(\frac{R^L}{p} +q_0 - \frac{R^L}{p_0}\right) \\
    &=
    pq_0
    +
    R^L\left(1-\frac{p}{p_0}\right)\\
    &\le
    pq_0
    +
    R_0\left(1-\frac{p}{p_0}\right)\\
    &=
    R_0,
\end{aligned}
\]
where we have used $R^L\le R_0$, which is implied by $\bar{q}^L = \frac{R^L}{p_0}$. Thus, $p_0$ maximizes aggregate profits for prices $p \in [p^L,p_0]$.

Finally, for $p>p_0$, $\widehat D^L(p)=0$, and
\[
    p\widehat D^H(p)\le R^H.
\]
Rationalizability implies that a feasible $\mathbf{D}$ exists, and for that $\mathbf{D}$:
\[
    R^H = p^HD^H(p^H) \leq p^H\left(D^H(p^H)+D^L(p^H) \right)\le R_0,
\]
by $p_0$ being aggregate-optimal in the feasible $\mathbf{D}$. Hence no price above $p_0$ yields aggregate revenue above $R_0$.

Thus $\widehat D^L$ and $\widehat D^H$ form a valid rationalization of the
observed outcomes. By construction, the upper bound on $L$ and the lower bound on $H$ are attained, so the upper bound is sharp.

\paragraph{Step 4. Proving the expression in the statement.}

We now study $\min\left\{\frac{R^L}{p},\,\frac{R_0}{p}-q^H\right\}$. For that, define $f(p) = \frac{R^L}{p} - \left(\frac{R_0}{p}-q^H \right)$. First, $f(p^L)<0$. Indeed, for any feasible $\mathbf{D}$,

\[ R^L + p^Lq^H \leq p^L\left(D^L (p^L) + D^H(p^L)\right) \leq R_0, \]
where the first inequality follows from monotonicity of $D^H$, and the second by $p_0$ being optimal in the aggregate. By reorganizing the inequality, we obtain the result.
The inequality above also shows that $R^L \leq R_0$, so $f'(p) = \frac{R_0-R^L}{p^2} > 0$, and $f$ crosses zero at most once. We can therefore divide the analysis into two cases. If $f(p_0) \leq 0$, which translates to $R^L+p_0 q^H \leq R_0$, then $\min\left\{\frac{R^L}{p},\,\frac{R_0}{p}-q^H\right\}= \frac{R^L}{p}$, for all $p \in [p^L,p_0]$. It follows that \eqref{eq: upper-bound-cs-two-segments} can be written as:
\begin{equation*}
    \Delta CS
    \le
    \int_{p^L}^{p_0}
    \frac{R^L}{p}dp
    -
    q^H(p^H-p_0), 
\end{equation*}

obtaining the expression in the statement of the theorem for this case upon integration. 

Otherwise---i.e., if  $R^L+p_0 q^H > R_0$---, there exists a unique $\hat{p}$ such that $f(\hat{p})=0$. In particular: $\hat{p} = \frac{R_0-R^L}{q^H}.$  \eqref{eq: upper-bound-cs-two-segments} can then be written as:
\begin{equation*}
    \Delta CS
    \le
    \int_{p^L}^{\hat p}
    \frac{R^L}{p}dp + \int^{p_0}_{\hat{p}}  \left(\frac{R_0}{p}-q^H\right)dp
    -
    q^H(p^H-p_0).
\end{equation*}
Integration delivers the second expression in the braces in the statement of the theorem. \hfill $\blacksquare$

\medskip

\subsection*{Proof of Theorem~\ref{thm:lower-bound-general}}
For later use, it will be useful to define 
\begin{equation}
p_k^\star
=
\frac{
\displaystyle
\min\left\{
R_0,\,
\sum_{j=k+1}^h R_j^H
\right\}
}{
\displaystyle
\sum_{j=k+1}^h q_j^H
+
\min\left\{
q_0-\underline Q^L-\sum_{j=1}^h q_j^H,\,
\min_{1\le m\le k}
\left[
\frac{R_0}{p_m^H}
-\sum_{j=m}^h q_j^H
\right]
\right\}
},
\end{equation}
and note that $\hat{p}_k = p^\star_k]^H_k$. 

Consider any demands that rationalize the observations, and write, for $i=1,\ldots,\ell$ and $j=1,\ldots,h$
\[
s_i^L:=D_i^L(p_0),
\qquad
s_j^H:=D_j^H(p_0).
\]
Since aggregate demand at $p_0$ is $q_0$, $\sum_{i=1}^{\ell}s_i^L+\sum_{j=1}^h s_j^H=q_0.$

\paragraph{Step 1. Lower bound for gains.}
We first characterize the smallest consumer-surplus gain that can arise
in the low-price segments. Segment-level optimality implies
\begin{equation}
   s_i^L\le \frac{R_i^L}{p_0},
\qquad
s_j^H\le \frac{R_j^H}{p_0}. 
\label{eq:segment-optimality}
\end{equation}

Hence, using aggregation of demand:
\[
\sum_{i=1}^{\ell}s_i^L
\ge
q_0-\sum_{j=1}^h\frac{R_j^H}{p_0}.
\]

There are additional lower bounds implied by aggregate optimality at
the observed low prices. Fix $k\in\{1,\ldots,\ell\}$. At $p_k^L$,
monotonicity implies that every low-price segment $i\ge k$ demands at
least $q_i^L$, while every other segment demands at least its quantity
at $p_0$. Therefore, using aggregation again
\[
p^L_k\left(q_0+\sum_{i=k}^{\ell}(q_i^L-s_i^L)\right) \leq p^L_k \left(\sum^\ell_{i=1} D^L_i + \sum^h_{j=1} D^H_j \right) =  p_k^L D_0(p^L_k) \leq R_0,
\]
where the second inequality holds by feasibility of $\mathbf{D}$, which implies $p_0$ is optimal in the aggregate market. Hence
\[
\sum_{i=k}^{\ell}s_i^L
\ge
q_0+\sum_{i=k}^{\ell}q_i^L-\frac{R_0}{p_k^L}.
\]
It follows from the definition of $\underline Q^L$ in \eqref{eq:Qlower} and non-negativity of $s^L_i$'s that

\begin{equation}
\sum_{i=1}^{\ell}s_i^L\ge \underline Q^L.
\label{eq:minimal-mass-requirement}
\end{equation}

We now show how to minimize the gains from low price segments. Notice that those gains are bounded below as follows:
\[
\sum^\ell_{i=1}\int_{p_i^L}^{p_0}D_i^L(p)\,dp
\ge
\sum^\ell_{i=1}(p_0-p_i^L)s_i^L.
\]

We now minimize the lower-bound above subject to the minimal mass requirement in \eqref{eq:minimal-mass-requirement} and the maximal level per segment given by segment optimality in \eqref{eq:segment-optimality}. This suggests maximizing the weight to segments with the highest prices, which have the lowest required consumer surplus gain. Formally:
\[
s_i^{L,*}
=
\min\left\{
\frac{R_i^L}{p_0},
\left[
\underline Q^L-\sum_{m=i+1}^{\ell}\frac{R_m^L}{p_0}
\right]_+
\right\}.
\]
In words, starting from $\ell$, we allocate as much mass as possible to this segment at $p_0$, up until either all the mass that needs to be allocated is given or we reach the maximal demand compatible with optimality of $p^L_\ell$. We then move to $\ell-1$ and repeat the procedure inductively with the remaining mass. Clearly $\sum^\ell_{i=1} s^L_i = \underline{Q}^L$. 

 Using this allocation, we obtain a lower bound for the gain in the low-price segments:
\[
\sum^\ell_{i=1} \int^{p_0}_{p^L_i} D^L_i(p) dp \geq   \sum_{i=1}^{\ell}
(p_0-p_i^L)
\min\left\{
\frac{R_i^L}{p_0},
\left[
\underline Q^L-\sum_{a=i+1}^{\ell}\frac{R_a^L}{p_0}
\right]_+
\right\}.
\]

\paragraph{Step 2. Upper bound for losses.}
We next characterize the largest possible consumer-surplus loss in the
high-price segments. For $p\in[p_0,p_j^H]$, write
\[
D_j^H(p)=q_j^H+z_j(p),
\]
with $z_j(p)\ge0$, and set $z_j(p)=0$ for $p>p_j^H$. Monotonicity
implies that every $z_j$ is nonincreasing.

We have:

\begin{equation}
    \sum^h_{j=1} \int^{p^H_j}_{p_0} D^H(p) dp = \sum^h_{j=1} q^H_j (p^H_j-p_0) + \sum^h_{j=1}\int^{p^H_h}_{p_0}z_j(p) dp. 
    \label{eq:CSchange-H}
\end{equation}

Our goal is to maximize the second term. At $p_0$, at least $\underline Q^L$ units must be allocated to the
low-price segments. Therefore, for every $p\ge p_0$,
\[
\sum_{j=1}^h z_j(p)
\le
q_0-\underline Q^L-\sum_{j=1}^h q_j^H.
\]

Now fix $k\in\{0,\ldots,h-1\}$ and consider $p\in(p_k^H,p_{k+1}^H)$, where $p_0^H=p_0$. Optimality in segment $j$ implies $z_j(p) \le \frac{R_j^H}{p}-q_j^H.$ Summing over segments with high price, and recalling $z_j(p)=0$ for all $j\leq k$ by definition, yields
\[
\sum_{j=1}^h z_j(p)
\le
\frac{\sum_{j=k+1}^hR_j^H}{p}
-
\sum_{j=k+1}^h q_j^H.
\]

Moreover, for any $1\leq m \leq k$,
\[
 p^H_m \left(\sum^h_{j=m} q^H_j + \sum^h_{j=1} z_j(p) \right)\leq p\left( \sum^h_{j=1} q^H_j  + \sum_{j=1}^h z_j(p)\right)
\le
R_0,
\]
where the first inequality follows from $p^H_m<p$ and $q^H_j$ being all positive, and the second inequality follows from aggregate optimality of $p_0$. It follows that, for $p\in(p_k^H,p_{k+1}^H)$,
\begin{align*}
\sum_{j=1}^h z_j(p)
\le & \\
\min \biggr\{
q_0-&\underline Q^L-\sum_{j=1}^h q_j^H,\,
\min_{1\le m\le k}
\left[
\frac{R_0}{p_m^H}-\sum_{j=m}^h q_j^H
\right],\,
\frac{
\min\left\{R_0,\sum_{j=k+1}^hR_j^H\right\}
}{p}
-\sum_{j=k+1}^h q_j^H
\biggr\},
\end{align*}

where, as in the statement, the minimum over $r$ is interpreted as
$+\infty$ when $k=0$.

The first two terms are constant in $p$, while the third is strictly
decreasing. By the definition of $p_k^\star$ in
\eqref{eq:pstar-high}, the constant and decreasing terms are equal at
$p_k^\star$. Hence, after restricting the crossing to the interval
$(p_k^H,p_{k+1}^H)$, the maximal feasible extra demand is constant up to $\widehat p_k=\min\left\{p_{k+1}^H,\,
\max\{p_k^H,p_k^\star\}\right\}$, and thereafter equals the third term in the minimization above.

It follows that: 

\begin{align*}
\sum^h_{j=1} \int^{p^H_{k+1}}_{p^H_{k}} z_j(p) dp \leq &\min\left\{
q_0-\underline Q^L-\sum_{j=1}^h q_j^H,\,
\min_{1\le r\le k}
\left[
\frac{R_0}{p_r^H}-\sum_{j=r}^h q_j^H
\right]
\right\}
(\widehat p_k-p_k^H)
\\
&\qquad+
\min\left\{
R_0,\sum_{j=k+1}^hR_j^H
\right\}
\log\left(\frac{p_{k+1}^H}{\widehat p_k}\right)
\\
&\qquad-
\left(\sum_{j=k+1}^h q_j^H\right)
(p_{k+1}^H-\widehat p_k).
\end{align*}

Summing over $k=0,\ldots,h-1$ and plugging in \eqref{eq:CSchange-H}, and combining the minimum consumer surplus gain proved in the previous step shows that the expression on the right-hand-side of \eqref{eq:general-lower-bound} is a lower bound for $\Delta \CS$. It remains to establish that it is the highest possible lower-bound.

\paragraph{Step 3. Constructing $\mathbf{D}$ to achieve the lower bound.} Using the quantities
$s_i^{L,*}$ defined above, set
\[
\widehat D_i^L(p)
=
\begin{cases}
q_i^L,
& p\le p_i^L,\\[1mm]
s_i^{L,*},
& p_i^L<p\le p_0,\\[1mm]
0,
& p>p_0.
\end{cases}
\]
Rationalizability guarantees
\[
0 \leq s_i^{L,*}\le\frac{R_i^L}{p_0}\leq q_i^L,
\]
so these demands are nonincreasing, nonnegative, and $p_i^L$ is optimal in each
low-price segment. Moreover, they clearly attain the lower bound in gains obtained in Step 1.

We define 
\[
\widehat D_j^H(p)
=
\begin{cases}
q_j^H+z_j(p_0),
& p\le p_0,\\[1mm]
q_j^H+z_j(p),
& p_0<p\le p_j^H,\\[1mm]
0,
& p>p_j^H,
\end{cases}
\]
where we describe the construction of $z_j$ below. 

To define $z_j$, let, for $p \in (p^H_k,p^H_{k+1})$, 
\begin{align*}
M(p) := & \\ \min\biggr\{
q_0-&\underline Q^L-\sum_{j=1}^h q_j^H,\,
\min_{1\le m\le k}
\left[
\frac{R_0}{p_m^H}-\sum_{j=m}^h q_j^H
\right],\,
\frac{
\min\left\{R_0,\sum_{j=k+1}^hR_j^H\right\}
}{p}
-\sum_{j=k+1}^h q_j^H
\biggr\},
\end{align*}

The function $M$ is nonincreasing and
absolutely continuous. We can then denote its almost-everywhere derivative $M'$ and, because $M$ is the minimum between a constant and a decreasing function of $p$, for almost all prices:
\[
-M'(p) \leq \max\left\{0, \frac{\min \{R_0,\sum_{j:p^H_j\geq p }R^H_j\}}{p^2}\right\}
\le
\frac{\sum_{\{j:p_j^H\ge p\}}R_j^H}{p^2}.
\]

We now define $z_j$ for each high-price segment $j$ as
\[
z_j(p)
=
\int_p^{p_j^H}
[-M'(s)]
\frac{R_j^H}
{\sum_{\{k:p_k^H\ge s\}}R_k^H}
\,ds,
\qquad
p< p_j^H,
\]
and set $z_j(p)=0$ for $p\geq p_j^H$. Since all revenues are positive, the
denominator is positive throughout the range of integration.

Because $M$ is nonincreasing, $-M'\ge0$ almost everywhere. Hence each
$z_j$ is nonnegative and nonincreasing in $p$. It clearly follows that $\widehat D^H$ is nonnegative and nonincreasing. Because $z_k(p^H_j)=0$, price-quantity pairs are consistent. We now prove that $p^H_j$ is optimal. Because demand is zero for $p>p^H_j$, clearly no such price can be optimal. Similarly, no price $p<p_0$ can be optimal because demand is constant in that region. For $p_0\le p \le p_j^H$,
\[
\begin{aligned}
z_j(p)
&=
\int_p^{p_j^H}
[-M'(s)]
\frac{R_j^H}
{\sum_{\{a:p_a^H\ge s\}}R_a^H}
\,ds\\
&\le
\int_p^{p_j^H}\frac{R_j^H}{s^2}\,ds=
\frac{R_j^H}{p}
-\frac{R_j^H}{p_j^H} =
\frac{R_j^H}{p}-q_j^H.
\end{aligned}
\]

Reorganizing, we obtain $p \widehat{D}^H(p) \leq R^H_j$, and the proof is done. 
Finally, 
\[
\begin{aligned}
\sum_{j=1}^h z_j(p)
&=
\int_p^{p_h^H}
[-M'(s)]
\frac{
\sum_{\{j:p_j^H\ge s\}}R_j^H
}{
\sum_{\{a:p_a^H\ge s\}}R_a^H
}
\,ds\\
&=
\int_p^{p_h^H}[-M'(s)]\,ds =
M(p),
\end{aligned}
\]
where the last equality uses $M(p_h^H)=0$. So $\widehat{D}_j^H$'s achieve the upper bound in losses derived in Step 2.

It remains to prove consistency and optimality of $p_0$ in the aggregate demand. At $p_0$, $\sum^h_{j=1}\widehat{D}_j^H(p_0) = M(p_0) + \sum^h_{j=1} q^H_j = q_0 - \underline{Q}^L$. Then, by construction of $\widehat{D}^L$ we have $D_0 (p_0) = \sum^h_{j=1}\widehat{D}^H (p_0)+\sum^l_{i=1} \widehat{D}^L(p_0)=q_0$, guaranteeing consistency.

For optimality, start with $p>p_0$. There:

\begin{equation*}
    p \sum^h_{j=1} \widehat{D}^H(p) = p \left(\sum_{j: p^k_j\geq p} q^H_j + \sum^n_{j=1} z_j(p) \right) \leq p\left(\sum_{j: p^k_j\geq p} q^H_j+\frac{R_0}{p}-\sum_{j:p^H_j\leq p} q^H_j\right),
\end{equation*}
where the inequality follows from $\sum_j z_j = M$, and from the definition of $M$. Therefore:
\[
p\sum_j\widehat D_j^H(p)\le R_0.
\]
For $p<p_0$, aggregate demand is constant between consecutive observed
low prices, so the only candidate optimal prices other than $p_0$ are $p^L_i$, $i=1,...,\ell$. At $p_k^L$:
\begin{align*}
p_k^L D_0 (p^L_k) &= p^L_k \left(q_0+\sum_{i=k}^{\ell}(q_i^L-s_i^{L,*})\right) \\ & \leq  p^L_k \left(q_0+\sum_{i=k}^{\ell}q_i^L\right) - p^L_k \underline{Q}^L \leq p^L_k \left(q_0+\sum_{i=k}^{\ell}q_i^L\right) - p^L_k \left(q_0 + \sum^\ell_{i=k} q_i^L - \frac{R_0}{p^L_k}\right) \leq R_0 ,    
\end{align*}

implying $p_0$ is optimal for the aggregate market.

The constructed demands therefore rationalize the observations and attain simultaneously the minimum low-side gain and the maximum high-side loss in \eqref{eq:general-lower-bound}. Hence the lower bound is sharp. \hfill $\blacksquare$

\subsection*{Proof of Theorem~\ref{thm:upper-bound-general}}

For later use, define 

\begin{equation}
 p^\star_k=\frac{
R_0-\sum_{i=1}^kR_i^L
}{
q_0-\overline Q^L+\sum_{i=k+1}^{\ell}q_i^L
}   ,
\label{eq:pstar-upper}
\end{equation}
and note $\tilde{p}_k = p^\star_k]^L_{k}.$

Consider any demands that rationalize the observations, and let
\[
s_i^L=D_i^L(p_0),
\qquad
s_j^H=D_j^H(p_0).
\]
Since aggregate demand at $p_0$ is $q_0$,
\[
\sum_{i=1}^{\ell}s_i^L+\sum_{j=1}^h s_j^H=q_0.
\]

\paragraph{Step 1. Upper bound for gains. }
Segment-level optimality implies
\[
s_i^L\le \frac{R_i^L}{p_0},
\]
while monotonicity implies $s_j^H\ge q_j^H$. Therefore, using the inequality above and the expression for aggregate demand:
\[
\sum_{i=1}^{\ell}s_i^L
\le
\min\left\{
\sum_{i=1}^{\ell}\frac{R_i^L}{p_0},
q_0-\sum_{j=1}^h q_j^H
\right\}
=
\overline Q^L.
\]

We now bound the gain in consumer surplus in the low-price segments.
For $k=1,\ldots,\ell$, consider a price $p\in(p_k^L,p_{k+1}^L)$, where $p_{\ell+1}^L:=p_0$. Summing segment-level optimality for $i=1,...,k$:
\[
\sum_{i=1}^k D_i^L(p)
\le
\frac{\sum_{i=1}^kR_i^L}{p}.
\]

Moreover, for that same p

\begin{align*}
R_0 &\geq p \left(\sum^k_{i=1} D^L_i(p) + \sum^\ell_{i=k+1} D^L_i (p) + \sum^h_{j=1} D^H_j (p) \right) \\ &\geq p \left(\sum^k_{i=1} D^L_i(p) + \sum^\ell_{i=k+1} q^L_i + q_0 - \overline{Q}^L \right),    
\end{align*}
where the first inequality follows from optimality of $p_0$ in the aggregate, and the second inequality uses the fact that $D^L_i$ is nonincreasing and $p < p^L_{k+1}$; and that $D^H_j$ is nonincreasing and, by aggregation $\sum^h_{j=1} D^H(p_0) \geq q_0 - \overline{Q}^L $.

Consequently,
\begin{equation}
\label{eq:upper-low-envelope}
\sum_{i=1}^kD_i^L(p)
\le
\min\left\{
\frac{\sum_{i=1}^kR_i^L}{p},
\frac{R_0}{p}
-
q_0+\overline Q^L
-
\sum_{i=k+1}^{\ell}q_i^L
\right\}.
\end{equation}

The two terms on the right-hand side of
\eqref{eq:upper-low-envelope} are equal at $p_k^\star$ defined in
\eqref{eq:pstar-upper}. The segment-level constraint is the smaller
one below $p_k^\star$, while the aggregate constraint is the smaller
one above it. Thus, after restricting the crossing to
$[p_k^L,p_{k+1}^L]$, the first constraint binds up to $\tilde{p}_k$
and the second thereafter. It follows that
\[
\begin{aligned}
\int_{p_k^L}^{p_{k+1}^L}
\sum_{i=1}^kD_i^L(p)\,dp
\le{}&
\left(\sum_{i=1}^kR_i^L\right)
\log\left(\frac{\tilde{p}_k}{p_k^L}\right)
\\
&+
R_0\log\left(\frac{p_{k+1}^L}{\tilde{p}_k}\right)
\\
&-
\left(
q_0-\overline Q^L
+\sum_{i=k+1}^{\ell}q_i^L
\right)
(p_{k+1}^L-\tilde{p}_k).
\end{aligned}
\]
Since
\[
\sum_{i=1}^{\ell}
\int_{p_i^L}^{p_0}D_i^L(p)\,dp
=
\sum_{k=1}^{\ell}
\int_{p_k^L}^{p_{k+1}^L}
\sum_{i=1}^kD_i^L(p)\,dp,
\]
summing these inequalities gives the first part of
\eqref{eq:general-upper-bound}.

\paragraph{Step 2. Lower bound for losses.} For a high-price segment $j$, monotonicity implies
\[
D_j^H(p)\ge q_j^H
\qquad
\text{for }p\in[p_0,p_j^H].
\]
Hence
\[
\int_{p_0}^{p_j^H}D_j^H(p)\,dp
\ge
q_j^H(p_j^H-p_0).
\]
Summing over $j$ gives the last term in
\eqref{eq:general-upper-bound}. We have therefore established the
upper bound. Combining with the lower bound in the previous step, we learn that the right-hand-side in \eqref{eq:general-upper-bound} is an upper bound to the change in consumer surplus. It remains to show that the bound is sharp. 

\paragraph{Step 3. Constructing $\mathbf{D}$ to achieve the upper bound. }We first construct the
low-price demands so that equality holds in
\eqref{eq:upper-low-envelope}. For
$p \in (p_k^L,p_{k+1}^L)$, define
\[
M(p) := \min\left\{
\frac{\sum_{i=1}^kR_i^L}{p},
\frac{R_0}{p}
-
q_0+\overline Q^L
-
\sum_{i=k+1}^{\ell}q_i^L
\right\}.
\]

Define:
\[
\widehat D_i^L(p)
= \begin{cases}
q^L_i, \qquad \text{if } p \leq p^L_i \\\min\left\{
\frac{R_i^L}{p},
\left[
M(p)
-
\sum_{a=i+1}^k\frac{R_a^L}{p}
\right]_+
\right\},  \qquad \text{if } p \in (p^L_k,p^L_{k+1}], \ k\geq i \\
0,  \qquad \text{if } p>p_0.
\end{cases}
\]

Clearly, $\hat{D}^L_i$ is nonnegative. We now prove it is non-increasing. Fix $i\le k$. On $(p_k^L,p_{k+1}^L]$,
\[
\begin{aligned}
\left[
M(p)-\sum_{a=i+1}^k\frac{R_a^L}{p}
\right]_+
=
\min\Bigg\{&
\frac{\sum_{a=1}^iR_a^L}{p},\\
&
\left[
\frac{R_0-\sum_{a=i+1}^kR_a^L}{p}
-\left(q_0-\overline Q^L
+\sum_{a=k+1}^{\ell}q_a^L\right)
\right]_+
\Bigg\}.
\end{aligned}
\]
The first term is decreasing in $p$. If
$R_0-\sum_{a=i+1}^kR_a^L\ge0$, the second term is also
nonincreasing; if this coefficient is negative, the expression inside the bracket is negative, so the second term is
identically zero. Hence the right-hand side is nonincreasing.
Since $R_i^L/p$ is also decreasing, $\widehat D_i^L(p)$ is
nonincreasing on $(p_k^L,p_{k+1}^L]$. Moreover, when $k$ grows only negative terms are added inside the minimum, so the whole expression in the second line is decreasing. 

Finally, to prove optimality of $p^L_i$ in the $i$-th low segment, note that demand is constant for $p\leq p^L_i$ and $0$ for $p>p_0$, so the only candidate prices other than $p_i^L$ are in $[p^L_i, p_0]$. Take one such price and let $k$ be such that $p^L_k \leq p \leq p^L_{k+1}$. Then:

$$p\widehat{D}^L_i(p) \leq p \frac{R^L_i}{p}, $$
proving optimality of $p^L_i$. 

We next construct demands for the segments in $H$. At $p_0$, the total low-price demand generated by this construction is
\[
\sum^\ell_{i=1} \widehat D^L(p_0) = \min\left\{
\sum_{i=1}^{\ell}\frac{R_i^L}{p_0},
q_0-(q_0-\overline Q^L)
\right\}
=
\overline Q^L.
\]
To construct the high-price demands, choose numbers $s_j^H$ satisfying
\[
q_j^H\le s_j^H\le\frac{R_j^H}{p_0}
\]
and
\[
\sum_{j=1}^h s_j^H=q_0-\overline Q^L.
\]
Such a choice exists because $ \sum_{j=1}^h q_j^H
\le
q_0-\overline Q^L
\le
\sum_{j=1}^h\frac{R_j^H}{p_0}.$ Indeed, the first inequality follows directly from $\overline Q^L
\le
q_0-\sum_{j=1}^h q_j^H$, by the definition of $\overline Q^L$. For the second, if
\[
\overline Q^L
=
q_0-\sum_{j=1}^h q_j^H,
\]
then
\[
q_0-\overline Q^L
=
\sum_{j=1}^h q_j^H
\le
\sum_{j=1}^h\frac{R_j^H}{p_0}.
\]
If instead
\[
\overline Q^L
=
\sum_{i=1}^{\ell}\frac{R_i^L}{p_0},
\]
then rationalizability implies
\[
R_0\le
\sum_{i=1}^{\ell}R_i^L+\sum_{j=1}^hR_j^H,
\]
and hence
\[
q_0-\overline Q^L
\le
\sum_{j=1}^h\frac{R_j^H}{p_0}.
\]

For each high-price segment, define
\[
\widehat D_j^H(p)
=
\begin{cases}
s_j^H,
& p\le p_0,\\[1mm]
q_j^H,
& p_0<p\le p_j^H,\\[1mm]
0,
& p>p_j^H.
\end{cases}
\]
These demands are nonincreasing. Moreover, for $p\le p_0$,
\[
p\widehat D_j^H(p)
\le
p_0s_j^H
\le R_j^H,
\]
while for $p_0<p\le p_j^H$,
\[
p\widehat D_j^H(p)
=
pq_j^H
\le
p_j^Hq_j^H
=
R_j^H.
\]
Thus $p_j^H$ is optimal in every high-price segment. By construction, the demands $\hat{D}^H_j$ hit the lower bound in Step 2.

It remains to show that the aggregate demand is consistent and that $p_0$ is optimal in the aggregate. For consistency,
\[
\widehat{D}_0(p_0) = \sum_{i=1}^{\ell}\widehat D_i^L(p_0)
+
\sum_{j=1}^h\widehat D_j^H(p_0)
=
\overline Q^L+(q_0-\overline Q^L)
=
q_0.
\]

For optimality, first note that demand is constant for any $p\leq p^L_1$, so no $p<p^L_1$ can be optimal. Take any
$p\in(p_k^L,p_{k+1}^L]$, the segments in $H$ have aggregate mass
$q_0-\overline Q^L$, the low-price segments $i>k$ each have mass
$q_i^L$, and the construction of the remaining low-price demands
gives
\[
\sum_{i=1}^k\widehat D_i^L(p)
\le
\frac{R_0}{p}
-
q_0+\overline Q^L
-
\sum_{i=k+1}^{\ell}q_i^L \leq \frac{R_0}{p},
\]
where we used that $q_0-\overline{Q}^L \geq \sum^h_{j=1} q^h_j\geq 0$ by the definition of $\overline{Q}^L$, and $\sum^\ell_{i=k+1} q^L_i \geq 0$. Thus, aggregate revenue is at most $R_0$ for every $p<p_0$.

For $p>p_0$, the low-price demands vanish. Between two consecutive
observed high prices, aggregate high-side demand is constant and equal
to the sum of the observed quantities of the segments whose prices
have not yet been reached. Rationalizability implies at every observed
high price $p_k^H$ that
\[
p_k^H\sum_{j=k}^h q_j^H\le R_0.
\]
Since revenue is increasing in price on each interval over which this aggregate demand is constant, aggregate revenue is therefore at most $R_0$ for every $p>p_0$ as well.

Thus the constructed demands rationalize the observations. They attain
equality in \eqref{eq:upper-low-envelope} at every price and satisfy
\[
\widehat D_j^H(p)=q_j^H
\qquad
\text{for }p\in(p_0,p_j^H].
\]
Hence they attain exactly the right-hand side of
\eqref{eq:general-upper-bound}, proving that the upper bound is sharp. \hfill $\blacksquare$

\newpage
\end{document}

%% file: Figures/worst-case-demand.tex
\begin{tikzpicture}
\begin{groupplot}[  
  tick label style={font=\footnotesize},
label style={font=\footnotesize},
title style={font=\footnotesize},
legend style={font=\footnotesize},
    group style={group size=2 by 1, horizontal sep=2.0cm},
    width=0.46\textwidth,
    height=0.34\textwidth,
    xmin=1.2, xmax=8.6,
    ymin=0, ymax=7,
    axis lines=left,
   xlabel={$p$},
xlabel style={
    at={(axis description cs:1,.1)},
    anchor=north west
},
    ylabel={$D(p)$},
    ylabel style={
    at={(axis description cs:0.1,1)},
    anchor=south east,
    rotate=270
},
    enlargelimits=false,
    clip=false
]

\nextgroupplot[
 xtick={2,5,6,8},
xticklabels={
    {\strut $p^L$},
    {\strut $p_0$},
    {\strut $\hat p$},
    {\strut $p^H$}
},
    ytick={2.3,3.8,5.5},
    yticklabels={$q_0-\overline q^H$,$q^L$,$q_0$}
]

\fill[gray!20] (axis cs:2,0) rectangle (axis cs:5,2.3);

\draw[densely dotted] (axis cs:2,0) -- (axis cs:2,3.8);

\addplot[very thick, color=orange] coordinates {
    (1.2,3.8)
    (2,3.8)
    (2,2.3)
    (5,2.3)
    (5,0)
    (8.6,0)
};

\addplot[only marks,mark=*,mark size=1.8pt, color=blue] coordinates {(2,3.8) (5,5.5) (8,2.4)};

\node at (axis cs:3.5,1.0) {$(+)$};

\nextgroupplot[
 xtick={2,5,6,8},
xticklabels={
    {\strut $p^L$},
    {\strut $p_0$},
    {\strut $\hat p$},
    {\strut $p^H$}
},
    ytick={2.4,3.2, 5.5},
    yticklabels={$q^H$,$\overline q^H$, $q_0$}
]

\fill[gray!20] (axis cs:5,0) rectangle (axis cs:6,3.2);

\addplot[name path=curveH, draw=none, domain=6:8, samples=150] {19.2/x};
\addplot[name path=axisH, draw=none, domain=6:8, samples=2] {0};
\addplot[gray!20] fill between[of=curveH and axisH];

\draw[densely dotted] (axis cs:5,0) -- (axis cs:5,3.2);
\draw[densely dotted] (axis cs:6,0) -- (axis cs:6,3.2);

\addplot[very thick, dashed, domain=3:6, samples=200, color=purple] {19.2/x};

\addplot[very thick, domain=1.2:6, samples=2, color=orange] {3.2};
\addplot[very thick, domain=6:8, samples=200, color=orange] {19.2/x};
\addplot[very thick, color=orange] coordinates {(8,2.4) (8,0)};

\addplot[only marks,mark=*,mark size=1.8pt, color=blue] coordinates {(2,3.8) (5,6.2) (8,2.4)};

\node at (axis cs:6.2,1.2) {$(-)$};

\end{groupplot}
\end{tikzpicture}

%% file: Figures/best-case-demand.tex
\begin{tikzpicture}
\begin{groupplot}[  
    tick label style={font=\footnotesize},
    label style={font=\footnotesize},
    title style={font=\footnotesize},
    legend style={font=\footnotesize},
    group style={group size=2 by 1, horizontal sep=2.0cm},
    width=0.46\textwidth,
    height=0.34\textwidth,
    xmin=1.2, xmax=8.6,
    ymin=0, ymax=4.5,
    axis lines=left,
    xlabel={$p$},
    xlabel style={
        at={(axis description cs:1,.1)},
        anchor=north west
    },
    ylabel={$D(p)$},
    ylabel style={
        at={(axis description cs:0.1,1)},
        anchor=south east,
        rotate=270
    },
    enlargelimits=false,
    clip=false
]

%
%
%
%
%


\nextgroupplot[
    xtick={2,4.111,5,8},
    xticklabels={
        {\strut $p^L$},
        {\strut $\widetilde p$},
        {\strut $p_0$},
        {\strut $p^H$}
    },
    ytick={1.2,3.0,3.8},
    yticklabels={
        {$\overline q^L$},
        {$q_0$},
        {$q^L$}
    }
]


\addplot[
    name path=lowcurveA,
    draw=none,
    domain=2:4.111,
    samples=200
] {7.6/x};

\addplot[
    name path=lowaxisA,
    draw=none,
    domain=2:4.111
] {0};

\addplot[gray!20]
fill between[of=lowcurveA and lowaxisA];


\addplot[
    name path=lowcurveB,
    draw=none,
    domain=4.111:5,
    samples=200
] {15/x-1.8};

\addplot[
    name path=lowaxisB,
    draw=none,
    domain=4.111:5
] {0};

\addplot[gray!20]
fill between[of=lowcurveB and lowaxisB];


\addplot[
    very thick,
    dashed,
    color=purple,
    domain=1.7:7,
    samples=200
] {7.6/x};

\addplot[
    very thick,
    dashed,
    color=gray,
    domain=2.5:7,
    samples=200
] {15/x-1.8};


\addplot[
    very thick,
    color=orange
] coordinates {
    (1.2,3.8)
    (2,3.8)
};


\addplot[
    very thick,
    color=orange
] coordinates {
    (1.2,3.8)
    (2,3.8)
};

\addplot[
    very thick,
    color=orange
] coordinates {
    (2,3.8)
    (2,3.8)
};

\addplot[
    very thick,
    color=orange,
    domain=2:4.111,
    samples=200
] {7.6/x};

\addplot[
    very thick,
    color=orange,
    domain=4.111:5,
    samples=200
] {15/x-1.8};

\addplot[
    very thick,
    color=orange
] coordinates {
    (5,1.2)
    (5,0)
    (8.6,0)
};


\draw[densely dotted]
    (axis cs:2,0) -- (axis cs:2,3.8);

\draw[densely dotted]
    (axis cs:4.111,0) -- (axis cs:4.111,1.848);


\addplot[
    only marks,
    mark=*,
    mark size=1.8pt,
    color=blue
] coordinates {
    (2,3.8)
    (5,3.0)
    (8,1.8)
};

\node at (axis cs:3.4,0.9) {$(+)$};


\nextgroupplot[
    xtick={2,4.111,5,8},
    xticklabels={
        {\strut $p^L$},
        {\strut $\widetilde p$},
        {\strut $p_0$},
        {\strut $p^H$}
    },
    ytick={1.8,3.0,3.8},
    yticklabels={
        {$q^H$},
        {$q_0$},
        {$q^L$}
    }
]


\fill[gray!20]
    (axis cs:5,0)
    rectangle
    (axis cs:8,1.8);


\addplot[
    very thick,
    color=orange
] coordinates {
    (1.2,1.8)
    (8,1.8)
    (8,0)
    (8.6,0)
};


\draw[densely dotted]
    (axis cs:5,0) -- (axis cs:5,3.0);


\addplot[
    only marks,
    mark=*,
    mark size=1.8pt,
    color=blue
] coordinates {
    (2,3.8)
    (5,3.0)
    (8,1.8)
};

\node at (axis cs:6.5,0.8) {$(-)$};

\end{groupplot}
\end{tikzpicture}

%% file: references.bib
@article{brown1996testable,
  title={Testable restrictions on the equilibrium manifold},
  author={Brown, Donald J and Matzkin, Rosa L},
  journal={Econometrica: Journal of the Econometric Society},
  pages={1249--1262},
  year={1996},
  publisher={JSTOR}
}

@article{chassang2022robust,
  title={Robust screens for noncompetitive bidding in procurement auctions},
  author={Chassang, Sylvain and Kawai, Kei and Nakabayashi, Jun and Ortner, Juan},
  journal={Econometrica},
  volume={90},
  number={1},
  pages={315--346},
  year={2022},
  publisher={Wiley Online Library}
}

@article{kang2025robustness,
  title={Robustness measures for welfare analysis},
  author={Kang, Zi Yang and Vasserman, Shoshana},
  journal={American Economic Review},
  volume={115},
  number={8},
  pages={2449--2487},
  year={2025},
  publisher={American Economic Association 2014 Broadway, Suite 305, Nashville, TN 37203}
}

@article{afriat1967construction,
  author  = {Afriat, Sydney N.},
  title   = {The Construction of Utility Functions from Expenditure Data},
  journal = {International Economic Review},
  year    = {1967},
  volume  = {8},
  number  = {1},
  pages   = {67--77}
}

@book{robinson1933economics,
  author    = {Robinson, Joan},
  title     = {The Economics of Imperfect Competition},
  year      = {1933},
  publisher = {Macmillan},
  address   = {London}
}

@book{pigou1920economics,
  author    = {Pigou, Arthur Cecil},
  title     = {The Economics of Welfare},
  year      = {1920},
  publisher = {Macmillan},
  address   = {London}
}

@article{schmalensee1981output,
  author  = {Schmalensee, Richard},
  title   = {Output and Welfare Implications of Monopolistic Third-Degree Price Discrimination},
  journal = {American Economic Review},
  year    = {1981},
  volume  = {71},
  number  = {1},
  pages   = {242--247}
}

@article{carvajal2013revealed,
  title={Revealed preference tests of the Cournot model},
  author={Carvajal, Andr{\'e}s and Deb, Rahul and Fenske, James and Quah, John K-H},
  journal={Econometrica},
  volume={81},
  number={6},
  pages={2351--2379},
  year={2013},
  publisher={Wiley Online Library}
}

@article{haghpanah2022limits,
  title={The limits of multiproduct price discrimination},
  author={Haghpanah, Nima and Siegel, Ron},
  journal={American Economic Review: Insights},
  volume={4},
  number={4},
  pages={443--458},
  year={2022},
  publisher={American Economic Association 2014 Broadway, Suite 305, Nashville, TN 37203}
}

@article{haghpanah2023pareto,
  title={Pareto-improving segmentation of multiproduct markets},
  author={Haghpanah, Nima and Siegel, Ron},
  journal={Journal of Political Economy},
  volume={131},
  number={6},
  pages={1546--1575},
  year={2023},
  publisher={The University of Chicago Press Chicago, IL}
}

@article{kartik2026lemonade,
  title={Lemonade from lemons: Information design and adverse selection},
  author={Kartik, Navin and Zhong, Weijie},
  journal={Theoretical Economics},
  volume={21},
  number={1},
  pages={281--324},
  year={2026},
  publisher={Wiley Online Library}
}

@article{elliott2026market,
  title={Market segmentation through information},
  author={Elliott, Matthew and Galeotti, Andrea and Koh, Andrew and Li, Wenhao},
  journal={Review of Economic Studies},
  pages={rdag067},
  year={2026},
  publisher={Oxford University Press UK}
}

@techreport{farboodi2025good,
  title={Good data and bad data: The welfare effects of price discrimination},
  author={Farboodi, Maryam and Haghpanah, Nima and Shourideh, Ali},
  year={2025},
  institution={National Bureau of Economic Research}
}

@article{varian1985price,
  author  = {Varian, Hal R.},
  title   = {Price Discrimination and Social Welfare},
  journal = {American Economic Review},
  year    = {1985},
  volume  = {75},
  number  = {4},
  pages   = {870--875}
}

@article{shih1988general,
  author  = {Shih, Jun-ji and Mai, Chao-cheng and Liu, Jung-chao},
  title   = {A General Analysis of the Output Effect under Third-Degree Price Discrimination},
  journal = {Economic Journal},
  year    = {1988},
  volume  = {98},
  number  = {389},
  pages   = {149--158}
}

@article{schwartz1990third,
  author  = {Schwartz, Marius},
  title   = {Third-Degree Price Discrimination and Output: Generalizing a Welfare Result},
  journal = {American Economic Review},
  year    = {1990},
  volume  = {80},
  number  = {5},
  pages   = {1259--1262}
}

@article{cowan2007welfare,
  author  = {Cowan, Simon},
  title   = {The Welfare Effects of Third-Degree Price Discrimination with Nonlinear Demand Functions},
  journal = {RAND Journal of Economics},
  year    = {2007},
  volume  = {38},
  number  = {2},
  pages   = {419--428}
}

@article{aguirre2010monopoly,
  author  = {Aguirre, I{\~n}aki and Cowan, Simon and Vickers, John},
  title   = {Monopoly Price Discrimination and Demand Curvature},
  journal = {American Economic Review},
  year    = {2010},
  volume  = {100},
  number  = {4},
  pages   = {1601--1615},
  doi     = {10.1257/aer.100.4.1601}
}

@article{cowan2012third,
  author  = {Cowan, Simon},
  title   = {Third-Degree Price Discrimination and Consumer Surplus},
  journal = {Journal of Industrial Economics},
  year    = {2012},
  volume  = {60},
  number  = {2},
  pages   = {333--345},
  doi     = {10.1111/j.1467-6451.2012.00479.x}
}

@article{cowan2016welfare,
  author  = {Cowan, Simon},
  title   = {Welfare-Increasing Third-Degree Price Discrimination},
  journal = {RAND Journal of Economics},
  year    = {2016},
  volume  = {47},
  number  = {2},
  pages   = {326--340},
  doi     = {10.1111/1756-2171.12128}
}

@article{bergemann2015limits,
  author  = {Bergemann, Dirk and Brooks, Benjamin and Morris, Stephen},
  title   = {The Limits of Price Discrimination},
  journal = {American Economic Review},
  year    = {2015},
  volume  = {105},
  number  = {3},
  pages   = {921--957},
  doi     = {10.1257/aer.20130848}
}

@article{deb2023revealed,
  author  = {Deb, Rahul and Kitamura, Yuichi and Quah, John K.-H. and Stoye, J{\"o}rg},
  title   = {Revealed Price Preference: Theory and Empirical Analysis},
  journal = {Review of Economic Studies},
  year    = {2023},
  volume  = {90},
  number  = {2},
  pages   = {707--743}
}
